\documentclass[12pt,hidelinks]{article}

\usepackage{hyperref}
\usepackage{scicite}
\usepackage{graphicx}
\usepackage{amsmath}
\usepackage{upgreek}
\usepackage{caption}
\usepackage{multirow}
\usepackage{subcaption}
\usepackage{gensymb}
\usepackage{booktabs}
\usepackage{rotating}
\usepackage{array}
\usepackage{tabularx}

\usepackage{makecell}

\usepackage[export]{adjustbox}

\newcolumntype{Y}{>{\centering\arraybackslash}X}
\usepackage[dvipsnames]{xcolor}

\newcolumntype{C}{>{\centering\arraybackslash}m{2.5cm}}
\newcolumntype{M}[1]{>{\centering\arraybackslash}m{#1}}

\usepackage{times}

\newenvironment{sciabstract}{%
\begin{quote} \bf}
{\end{quote}}

\title{A van der Waals material exhibiting room temperature broken inversion symmetry with ferroelectricity}

\author
{Fabia F. Athena$^{1,2,\ast,\dagger}$, Cooper A. Voigt$^{3,\ast}$, Mengkun Tian$^{4}$, Anjan Goswami$^{1}$, \\
Emily Toph$^{3}$, Moses Nnaji$^{3}$, Fanuel Mammo$^{3}$, Brent K. Wagner$^{5}$,\\
Sungho Jeon$^{6}$, Wenshan Cai$^{1}$, Eric M. Vogel$^{1,3,\dagger}$\\
\\
\normalsize{$^{1}$School of Electrical and Computer Engineering, Georgia Institute of Technology}\\
\normalsize{Atlanta, GA, USA}\\
\normalsize{$^{2}$Department of Electrical Engineering, Stanford University, Stanford, CA, USA}\\
\normalsize{$^{3}$School of Materials Science and Engineering, Georgia Institute of Technology}\\
\normalsize{$^{4}$Institute for Matter and Systems, Georgia Institute of Technology, Atlanta, GA, USA}\\
\normalsize{$^{5}$Georgia Tech Research Institute, Atlanta, GA, USA}\\
\normalsize{$^{6}$Materials Science and Engineering, University of Pennsylvania, Philadelphia, PA, USA}\\
{\normalsize\textit{$^{\dagger}$E-mail: } \textit{eric.vogel@mse.gatech.edu, fathena@stanford.edu}, \textit{$^{\ast}$Equal contributions.}}
}

\date{}

\begin{document}

\baselineskip24pt

\maketitle

\begin{sciabstract}
Since the initial synthesis of van der Waals two-dimensional indium selenide was first documented in 1957, five distinct polymorphs and their corresponding polytypes have been identified. In this study, we report a unique phase of indium selenide via Scanning Transmission Electron Microscopy (STEM) analysis in the synthesized large-area films -- which we have named the $\upbeta^\text{p}$ phase. The quintuple layers of the $\upbeta^\text{p}$ phase, characterized by a unique zigzag atomic configuration with unequal indium-selenium bond lengths from the middle selenium atom, are distinct from any other previously reported phase of indium selenide. Cross-sectional STEM analysis has revealed that the $\upbeta^\text{p}$ layers exhibit intralayer shifting. We found that indium selenide films with $\upbeta^\text{p}$ layers display electric-field-induced switchable polarization characteristic of ferroelectric materials, suggesting the breaking of the inversion symmetry. Experimental observations of nonlinear optical phenomena -- Second Harmonic Generation (SHG) responses further support this conclusion. This study reports a $\upbeta^\text{p}$ phase of indium selenide showing ferroelectricity over large areas at room temperature in a low-dimensional limit.
\end{sciabstract}

\section{Introduction}

Low-dimensional materials hold significant promise for high-density logic and memory devices, sensors, and actuators, owing to their ability to maintain functional properties even within a few atomic layers \cite{damjanovic2001ferroelectric,muralt2000ferroelectric,george2016nonvolatile,yu2017flexible}. The absence of inversion symmetry in noncentrosymmetric 2D crystal structures leads to intriguing phenomena such as ferroelectricity, which has potential for emerging high-density logic and memory applications \cite{shi2016symmetry,kenzelmann2005magnetic,sai2000compositional}. Consequently, it has been widely believed that materials with centrosymmetric configurations cannot exhibit ferroelectric properties \cite{PhysRevLett.99.177202,li2023emergence,zhang2019toward}. However, recent research has confirmed the presence of room-temperature ferroelectricity arising from intralayer sliding in 2D van der Waals (vdW) GaSe monolayers, which were previously assumed to be centrosymmetric \cite{li2023emergence}. This discovery suggests that it is worthwhile to investigate other nominally centrosymmetric vdW 2D materials, as they may similarly exhibit unexpected noncentrosymmetric behaviors, opening new avenues for further exploration \cite{cohen1992origin,valasek1921piezo,zhang2023ferroelectric}.

VdW layered 2D materials are also notable for their rich polymorphism, whereby subtle variations in interlayer and intralayer stacking of the quintuple layers yield multiple crystalline phases \cite{choi2017electrically, tan2023polymorphism,zhou2017out,zhou2015controlled}. Among these materials, 2D vdW indium selenides (In$_2$Se$_3$, InSe) are particularly promising due to their rich physical properties. To date, five distinct phases have been identified—$\upalpha$, $\upbeta$, $\upgamma$, $\upkappa$, and $\updelta$ \cite{ding2017prediction}. Notably, certain noncentrosymmetric phases mitigate thickness-related depolarization effects commonly encountered in conventional ferroelectrics \cite{zheng2018room}. The noncentrosymmetric $\upalpha$-phase of In$_2$Se$_3$ has gained significant interest due to its potential as a ferroelectric semiconductor, capable of maintaining ferroelectricity in both in-plane and out-of-plane directions—even at the monolayer limit \cite{cui2018intercorrelated,xue2018room}. Researchers have also demonstrated the coexistence of out-of-plane and in-plane ferroelectricity in $\upbeta$-InSe nanoflakes at room temperature, driven by intralayer sliding of selenium atomic sublayers, which disrupts local symmetry \cite{hu2019room}. Additional studies on $\upbeta'$-In$_2$Se$_3$ flakes reported robust in-plane ferroelectricity at temperatures up to 200 $\degree$C in both bulk and exfoliated layers \cite{zheng2018room}. Recent investigations further revealed ferroelectric switching in $\upbeta'$-In$_2$Se$_3$ resulting from tip-induced flexoelectric effects and ferroelastic transitions between the $\upbeta'$ and $\upbeta''$ phases. These findings provide valuable insights into the complex mechanisms and atomic-scale configurations that govern ferroelectricity in different phases of indium selenide \cite{chen2021atomic}.

Advancements in deposition techniques such as Molecular Beam Epitaxy (MBE) \cite{joyce1985molecular} and Chemical Vapor Deposition (CVD) have enabled the synthesis of different phases of large-area 2D $\upbeta$-, $\upbeta'$-, and $\upalpha$ indium selenide films, marking significant progress in the scalable production of 2D ferroelectric semiconductors \cite{Han2022PhasecontrollableLT}. These developments not only support the growth of centimeter-scale 2D indium selenide films across critical paraelectric ($\upbeta$), ferroelectric ($\upalpha$), and antiferroelectric ($\upbeta'$) phases, but also provide viable strategies for phase transition control. This, in turn, enhances the scalable fabrication of 2D thin-film-based memories and heterophase junctions with improved non-volatile memory functionality \cite{han2023phase}. Collectively, these advances help address the memory-wall challenge in data-intensive computing for artificial intelligence (AI), paving the way for high-density memory technologies compatible with N3XT (Nano-Engineered Computing Systems Technology) 3D integration \cite{franklin2022carbon,athena2024first}.

In this study, we report a unique quintuple layer of indium selenide, which is distinct from any previously reported phase of In$_2$Se$_3$ and InSe, in a large-area continuous film deposited using MBE. We have designated this new phase the $\upbeta^\text{p}$ phase. The $\upbeta^\text{p}$ phase is marked by a unique zigzag atomic configuration and unequal indium-selenium bond lengths within the quintuple layer. Our cross-sectional STEM analysis revealed intralayer shifting in the $\upbeta^\text{p}$ phase, reminiscent of polarization mechanisms in noncentrosymmetric 2D systems. Notably, these films demonstrate an electric-field-tunable polarization. Experimental investigations have detected nonlinear optical phenomena, such as SHG, thus confirming the breaking of inversion symmetry. The electric‑field‑induced enhancement of the SHG signal, together with the bias‑driven increase of the $\upbeta^{\text{p}}$ phase observed by STEM, indicates a bias-induced phase transition. Our findings introduce the $\upbeta^\text{p}$ phase of indium selenide as a noncentrosymmetric low-dimensional material that exhibits ferroelectricity in large-area thin films at room temperature.

\section{Results and Discussion}

\begin{figure}[]
    \centering
    \includegraphics[width=\textwidth]{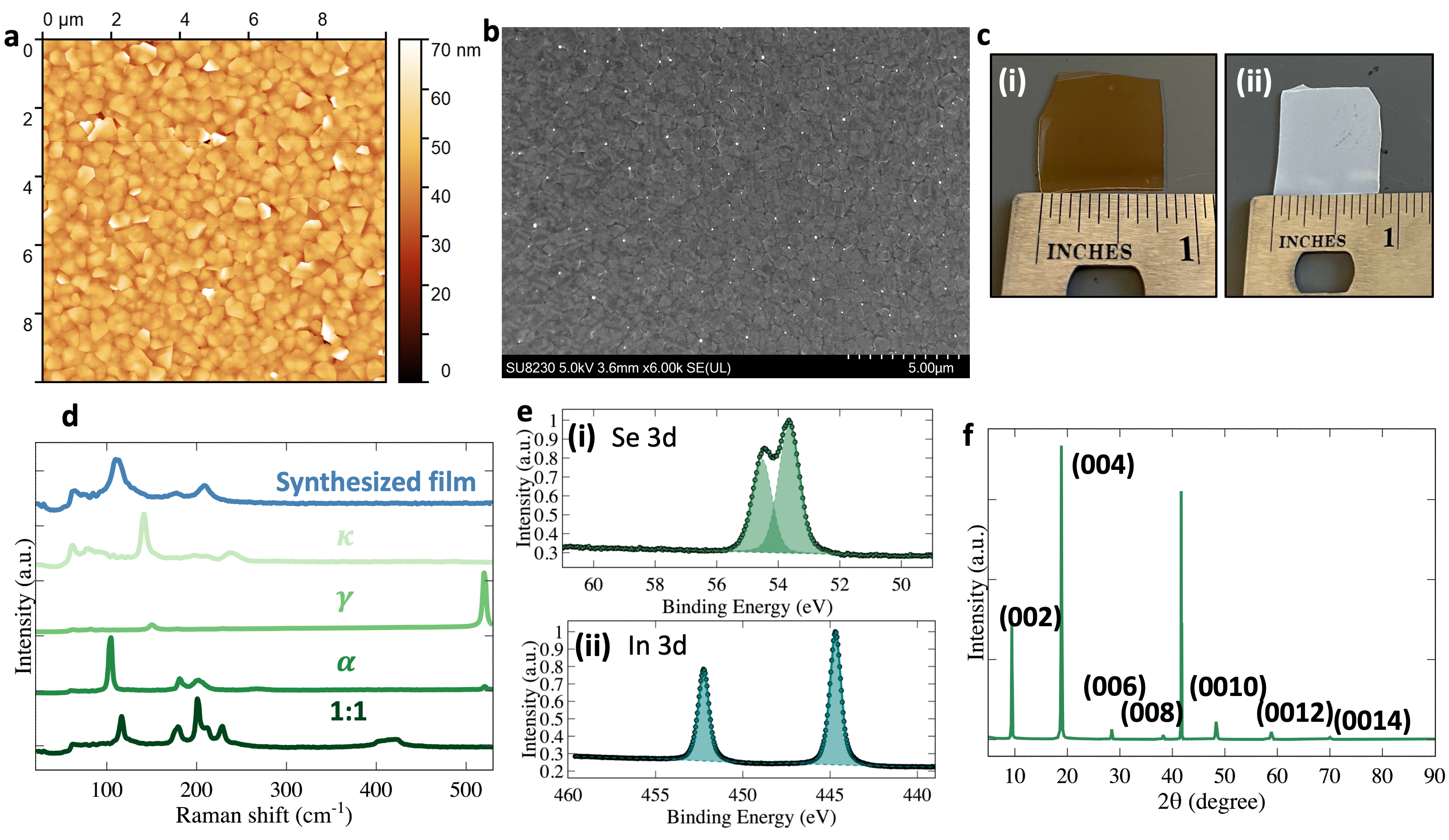}
    \caption{Characterization of synthesized large area indium selenide thin films. (a) AFM and (b) SEM image. (c) (i) and (ii) illustrate the optical images showing the synthesized large area $\upbeta$-indium selenide thin film and the bare sapphire substrate, respectively, demonstrating the extensive coverage. (d) Raman spectra comparing the synthesized film to various other phases, including $\upalpha$, $\upgamma$, and $\upkappa$ phases. The synthesized indium selenide film exhibits a distinct Raman peak position that is characteristic of the $\upbeta$ phase, differentiating it from other investigated phases. (e) XPS analysis. The Se/In stoichiometry, determined using Scofield RSFs (In/Se RSF ratio = 9.8), is $\sim$1.4. (f) XRD characterization reveals the synthesized indium selenide film primarily consists of the $\upbeta$ phase, as evidenced by corresponding (002), (004), (006), (008), (0010), (0012), and (0014) peaks. 
}
    \label{fig:1}
\end{figure}

Large-area indium selenide films were synthesized on c-plane sapphire substrates using MBE. The growth conditions, including a Se/In beam-equivalent-pressure ratio of $\sim$15 and a substrate temperature of $\sim$700~$\degree$C, are provided in Section~\ref{sec:app:mbe} and Figure~\ref{fig:s1}. The synthesized indium selenide films were characterized to evaluate properties such as homogeneity, surface morphology, and chemical composition, as illustrated in Figure~\ref{fig:1}. Figures~\ref{fig:1}(a) and \ref{fig:1}(b) show Atomic Force Microscopy (AFM) and Scanning Electron Microscopy (SEM) images, respectively. The AFM height map reveals sub-micron scale grains with a root-mean-square roughness of 0.4 nm, a level sufficiently low to prevent roughness-induced degradation of mobility and on-current~\cite{geiger2020effect}. Additionally, the SEM image confirms a continuous, pinhole-free film extending over an area coverage of $>$0.5 cm $\times$ 0.5 cm. Optical images in Figure~\ref{fig:1}(c)(i) and (ii) show the synthesized indium selenide thin film (brown color) and the bare sapphire substrate (white color), respectively, further demonstrating extensive coverage and uniformity. Raman spectra in Figure~\ref{fig:1}(d) compares the synthesized film with reference films such as indium monoselenide (InSe) and other phases of In$_2$Se$_3$, including the $\upalpha$, $\upgamma$, and $\upkappa$ polymorphs. The synthesized film (blue trace) shows Raman peaks at 110 cm$^{-1}$, 180 cm$^{-1}$, and 205 cm$^{-1}$. X-ray photoelectron spectroscopy (XPS) characterization performed to analyze the chemical composition, shown in Figure~\ref{fig:1}(e), reveals that the nominal Se/In stoichiometry (using Scofield RSFs, In/Se RSF ratio = 9.8) of the film is approximately 1.4. Moreover, X-ray diffraction (XRD) analysis was performed to reveal the crystal structure of the synthesized film. The XRD peaks closely resemble the crystal planes of the $\upbeta$-phase, including (002), (004), (006), (008), (0010), (0012) and (0014) \cite{tan2023polymorphism}, indicating that the synthesized thin film primarily consists of the $\upbeta$-In$_2$Se$_3$ phase, which is a centrosymmetric, non-ferroelectric phase of indium selenide.

Figure~\ref{fig:2}(a) provides a schematic representation of the top-gated field-effect transistor (FET) device structure fabricated from the synthesized film. The fabrication process flow is described in Figure~\ref{fig:s2}. An optical micrograph of a representative device is shown in Figure~\ref{fig:2}(b). Cross‑sectional energy‑dispersive X‑ray spectroscopy (EDS) mapping (Figure \ref{fig:2}(c)) confirms a uniform distribution of hafnium, titanium, gold, and aluminum in their respective layers, with indium and selenium localized in the channel of the device stack. No detectable intermixing between adjacent layers is observed, ensuring compositional integrity. The transfer characteristics (I\textsubscript{D} vs V\textsubscript{GS}) in Figure~\ref{fig:2}(d) exhibit an electric field-induced hysteresis loop in the clockwise direction, similar to previously reported ferroelectric In\textsubscript{2}Se\textsubscript{3} FeFETs~\cite{rodriguez2020electric,si2019ferroelectric,han2023phase,fei2018ferroelectric}. Details of the electrical testing and connections are provided in Figure~\ref{fig:s3}. The hysteresis loops are tunable with respect to the gate and drain biases, similar to typical FeFET characteristics. As the gate-voltage swing increases, the hysterisis window widens, confirming robust switchable polarization in the indium selenide layers. The calculated subthreshold slope (SS) is around 7.688 V/dec. The contact resistance was further measured using the transmission line method (TLM). Figure~\ref{fig:S4TLM}, shows that the contact resistance, R\textsubscript{C}, is within a reasonable range, approximately 2.86 k$\Omega$. The output characteristics (I\textsubscript{D} vs V\textsubscript{DS}), shown in Figure~\ref{fig:2}(f), measured from 0.0 V to 3.5 V, exhibit increasing current with gate bias, further demonstrating gate-controlled conductivity. Furthermore, sweep-rate dependency tests were performed to examine if the hysteresis loops change with sweep-rate variation. The hysteresis loops remain almost unchanged across varying sweep rates, ranging from 96 mV/s to 420 mV/s in increments of 15 mV/s for devices having two different dimensions, as depicted in Figures~\ref{fig:2}(g) and (h), respectively. The consistency of hysteresis loops across different sweep rates suggests that their origin is not trap-induced but rather stems from noncentrosymmetry- and polarization-induced~\cite{han2023phase,pinto2019effect}, origin. Moreover, transient current measurements were conducted by applying a constant bias and monitoring the temporal change in drain current, as detailed in Figure~\ref{fig:S5}. A high density of traps typically produces either a sharp or broad peak in current due to trapping and detrapping dynamics during transient current measurements~\cite{bronuzzi2019transient}. In the case of a baseline trap-rich indium selenide film grown on a SiO$_2$ substrate, a pronounced peak is observed. Additionally, in separate experiments, FET devices fabricated from these trap-rich films lack consistent, well-defined hysteresis (data not shown). In contrast, optimized indium selenide films grown on sapphire substrates exhibit no transient current peak and show reproducible, sweep-rate-independent hysteresis across multiple device batches. These set of experiments suggest that the switchable polarization originates from broken inversion symmetry. However, the synthesized film has been nominally identified as the $\upbeta$-In$_2$Se$_3$ phase by XRD measurements, which is centrosymmetric and non-ferroelectric. How can a centrosymmetric film yield polarization-induced hysteresis? To address this question, we performed SHG measurements to investigate the true symmetry of the synthesized films.

Figure~\ref{fig:2}(i) and (j) illustrate the nonlinear-optical response of the synthesized indium selenide films. An incident wavelength of 800~nm generates a strong SHG response at 400~nm, as shown in Figure~\ref{fig:2}(i), which indicates the absence of centrosymmetry in the film. Details of the SHG measurement setup and methodology are provided in Section~\ref{sec:app:shg} and Figure~\ref{fig:s6}. Further, an enhancement in the SHG response is observed with increasing incident laser power, as shown in Figure~\ref{fig:2}(j). Power levels ranging from 0.7~mW to 1.3~mW in 0.1~mW increments show a corresponding increase in SHG intensity. The inset of Figure~\ref{fig:2}(j) illustrates the relationship between second harmonic intensity (SHG intensity) and incident laser power on a log-log scale, revealing a linear correlation with a slope of 2. This quadratic dependence further confirms the robust SHG response, underscoring the broken inversion symmetry and ferroelectricity in the synthesized indium selenide film~\cite{janisch2014extraordinary,malard2013observation}.
Additionally, control measurements were conducted on blank sapphire substrates and amorphous $\upgamma$-phase In$_2$Se$_3$ films. Neither exhibited any SHG response across various locations, as shown in Figure~\ref{fig:S7}, which is expected and consistent with previous studies~\cite{cammarata2014microscopic,jing2018joint}. Furthermore, the reproducibility and reliability of the SHG response were verified by performing SHG measurements on optimized indium selenide films synthesized from different batches, as illustrated in Figure~\ref{fig:S8}. These observations confirm that the synthesized indium selenide films possess a noncentrosymmetric crystal structure with broken inversion symmetry.

To further understand the atomic-scale origin of the observed noncentrosymmetry, we conducted an in-depth STEM analysis. Figure~\ref{fig:3} presents detailed STEM imaging that reveals the microstructure of the synthesized indium selenide films. Figure~\ref{fig:3}(a) shows a top-view STEM image, clearly depicting a predominantly $\upbeta$-phase In$_2$Se$_3$ surface. A cross-sectional STEM image, shown in Figure~\ref{fig:3}(b), reveals that the film primarily comprises quintuple layers of the $\upbeta$-phase with a 3R polytype configuration which is centrosymmetric. Interestingly, some quintuple layers, which are vdW bonded to the $\upbeta$-phase, exhibit a unique zigzag atomic configuration. Since these embedded zigzag layers are encapsulated within the predominant $\upbeta$-phase layers, they remain undetected in top-view STEM imaging or XRD or Raman measurements of the pristine film. Furthermore, the structure and bond distances of this unique zigzag phase are investigated using high-angle annular dark-field (HAADF)-STEM imaging, as shown in Figure~\ref{fig:3}(c)--(g). The summed intensity for each atomic column in the $\upbeta$ and zigzag layers is shown in Figures~\ref{fig:3}(d) and \ref{fig:3}(e), respectively. Interatomic column distances are also calculated for both phases. The red and blue circles represent the centers of the In and Se atoms, respectively. For the $\upbeta$ phase as shown in Figure~\ref{fig:3}(f), the center In--Se and Se--In interatomic distance is measured as $0.228\pm 0.017$~ nm (yellow lines) and the Se--In interatomic distance is $0.171\pm 0.004$~nm (green line), which are close to In$_2$Se$_3$ reference values of $0.22$ nm and $0.175$ nm\cite{tan2023polymorphism}. In contrast, in the zigzag quintuple layers, as shown in Figure~\ref{fig:3}(g), the left Se--In interatomic distance is $0.151\pm 0.008$~nm, while the distance for the two right Se--In atomic columns is $0.132\pm 0.012$~nm. Here, STEM analysis in Figure 3(a)--(b) was performed using a Hitachi HD2700 aberration-corrected STEM, and STEM images in Figure 3(c)--(g) were analyzed using a NEOARM STEM to achieve better resolution. These bond distances and consequently the distinctive zigzag atomic configuration of the quintuple layers do not match those of any previously reported indium selenide phase (Table \ref{tab:phaselist}), to the best of our knowledge. We designate this phase as the $\upbeta^\text{p}$ phase. This unique atomic structure is the origin of the unexpected properties observed in the film, which we explore further under electrical biasing.

The effect of electrical biasing on the atomic configuration of the $\upbeta^\text{p}$ phase was further analyzed. For the biased film, a gate bias sweeping from 0 V to 2 V was applied with a drain bias of 1.5 V. It was observed that electrical biasing significantly increases the presence of the zigzag $\upbeta^\text{p}$ phase in the film, as shown in Figure~\ref{fig:4}. This observation suggests that electrical biasing can modify the atomic configuration within the film, increasing the proportion of the $\upbeta^\text{p}$ phase (cyan shaded region) compared to the $\upbeta$ phase (blue shaded region). Such alterations may impact the electronic properties, reflecting the bias-induced ferroelectricity previously reported in studies on In$_2$Se$_3$ and InSe. Two types of zigzag monolayers have been identified: type A and type B, the latter being a rotated version of type A. Furthermore, a statistical quantitative analysis of the $\upbeta$ and $\upbeta^\text{p}$ phases in biased and pristine films was performed on STEM images. A significant increase in the proportion of the $\upbeta^\text{p}$ phase was observed in electrically biased films compared to pristine films, rising from 0.83\% to 13.6\%, as detailed in Table~\ref{tab:1}. An approximately 16-fold increase in the $\upbeta^\text{p}$ phase in the biased film is observed. Details of the calculation are provided in Table~\ref{tab:S2}.

The inset of Figure \ref{fig:4}(c) shows a magnified view of a $\upbeta^{\text{p}}$ quintuple layer of $\upbeta^\text{p}$ phase. This layer features a distinctive zigzag arrangement of atomic columns and unequal In–Se bond lengths. A lateral displacement of the indium sublayer, indicated by the red arrow, breaks the local mirror symmetry and gives rise to the ferroelectric response. Comparable intralayer sliding has been reported to induce ferroelectricity in otherwise non‑ferroelectric GaSe \cite{li2023emergence}. In indium selenides, shifted or zigzag stacking motifs within or between quintuple layers underpin the ferroelectric behaviour of the $\upalpha$ and $\upbeta'$ phases, where broken inversion symmetry leads to spontaneous polarization. The same zigzag configuration with shifted atomic sublayer in $\upbeta^{\text{p}}$ phase therefore provides the structural origin of the observed second‑harmonic‑generation response and electric‑field‑switchable polarization in our synthesized indium selenide film.

Figure~\ref{fig:4}(d) shows three vdW bonded quintuple layers in a biased film, where a smooth transition from the centrosymmetric $\upbeta$ phase to the distinct zigzag $\upbeta^\text{p}$ phase is observed in the third layer. This transition illustrates the evolution of the noncentrosymmetric $\upbeta^\text{p}$ phase from the centrosymmetric and non-ferroelectric $\upbeta$ phase through atomic displacements, underscoring the intricacy and potential of bias and stress-induced phase transformations in vdW-bonded materials. Figure~\ref{fig:4}(e) compares the Raman spectra of pristine and electrically biased films. The subtle shift in the Raman spectra upon applying an electric bias suggests slight modifications in lattice dynamics, due to an increased proportion of the $\upbeta^\text{p}$ phase within the biased film, although the change is minimal. Additionally, Figure~\ref{fig:4}(f) presents the analysis of the nonlinear optical responses of pristine and biased films. An increase in the SHG peak intensity (approximately 1.5 times) is observed in the electrically biased film compared to the pristine film, further confirming that the increase of the $\upbeta^\text{p}$ phase under electric biasing. This finding aligns with the results from STEM analysis, indicating an electric bias-induced phase transition from the $\upbeta$ phase to the $\upbeta^\text{p}$ phase. These insights highlight the versatility and potential for modulating the properties of the synthesized indium selenide film through electrical biasing, opening avenues for innovative applications in electronics and photonics.

\section{Conclusion}

This study reports $\upbeta^\text{p}$ phase of indium selenide, distinguished by a unique zigzag atomic configuration and unequal indium–selenium bond lengths within its quintuple layers. The phase originates from atomic displacements—most notably a lateral shift of the central Se sublayer—that break inversion symmetry. Comprehensive analyses—including STEM, SHG measurements, and electrical characterizations— show that the $\upbeta^\text{p}$ phase possesses broken inversion symmetry and exhibits room-temperature switchable polarization across large areas. Notably, under electric bias the fraction of the $\upbeta^\text{p}$ phase increases by approximately 16 times, underscoring the potential for electric-field driven phase modulation. Further investigation of this room-temperature, ferroelectric van der Waals $\upbeta^{\text{p}}$ phase could unlock both electronic and photonic applications—especially scalable, high-density, 3D-integrated non-volatile memories for low-power AI based on low-dimensional materials.

\begin{figure}[]
    \centering
    \includegraphics[width=0.85\textwidth]{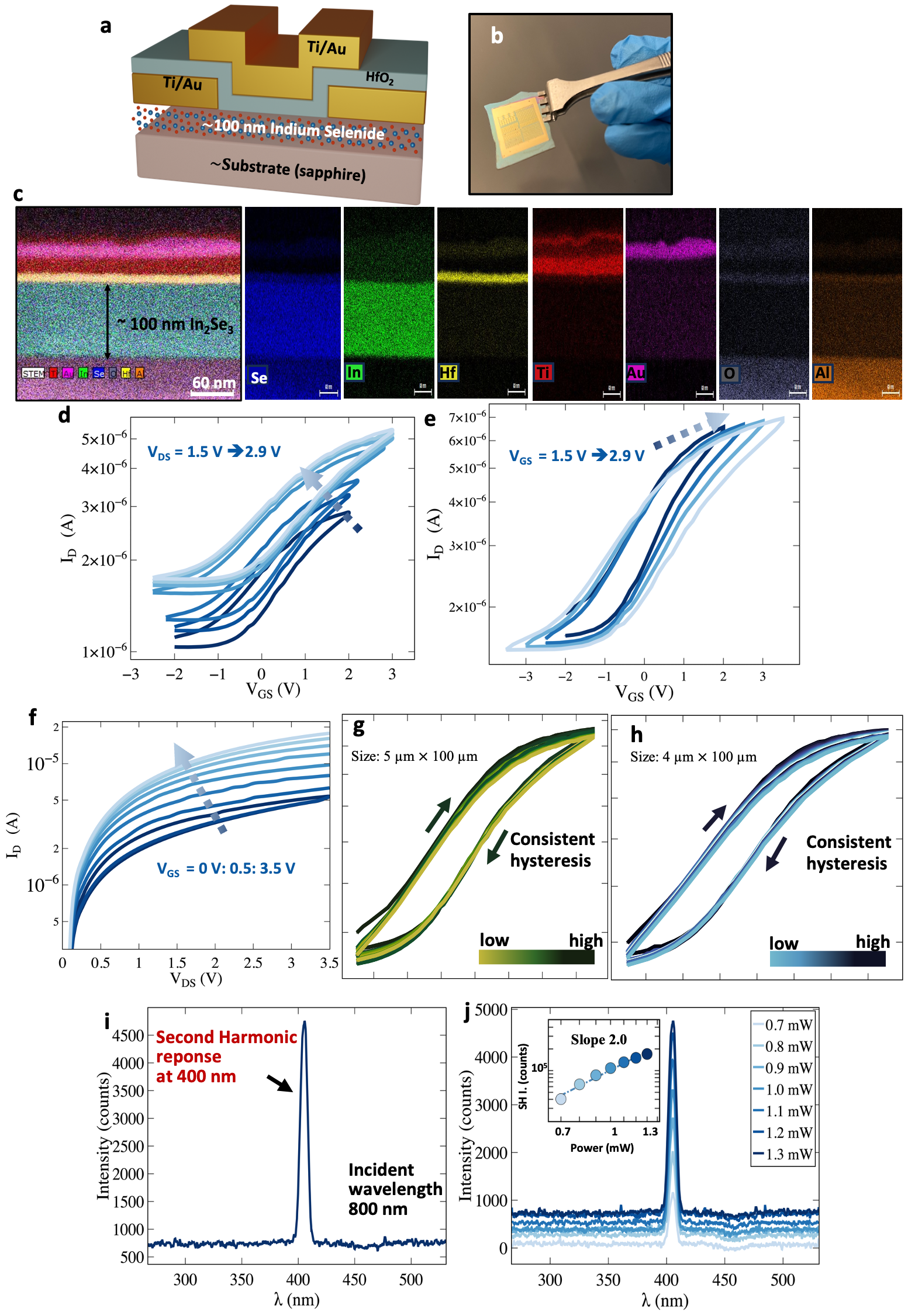}
\end{figure}

\begin{figure}[]
    \centering
    \caption{Characterization of FETs fabricated using the synthesized indium selenide film as the channel material. (a) Schematic of the device structure. (b) Optical image. (c) Energy-dispersive x-ray spectroscopy mapping of the device cross-section. (d) Transfer characteristics illustrate electric field-induced clockwise hysteresis loops. (e) An increase in the hysteresis loop size with an increase in gate bias indicates tunable switching window. (f) Output characteristics. Hysteresis loops at various sweep rates for different device sizes (g) 5 $\upmu$m $\times$ 100 $\upmu$m and (h) 4 $\upmu$m $\times$ 100 $\upmu$m. The loops maintain consistency at different sweep rates from 96 mV/s to 420 mV/s with an increment of 15 mV/s, indicative of polarization-induced hysteresis. (i) SHG response observed at 400 nm wavelength with an incident wavelength of 800 nm. (j) The SHG response increases with the power of the incident laser; The inset intensity-versus-power shows a slope of 2, further confirming the noncentrosymmetry.
}
    \label{fig:2}
\end{figure}

\begin{figure}[h]
    \centering
    \includegraphics[width=1\textwidth]{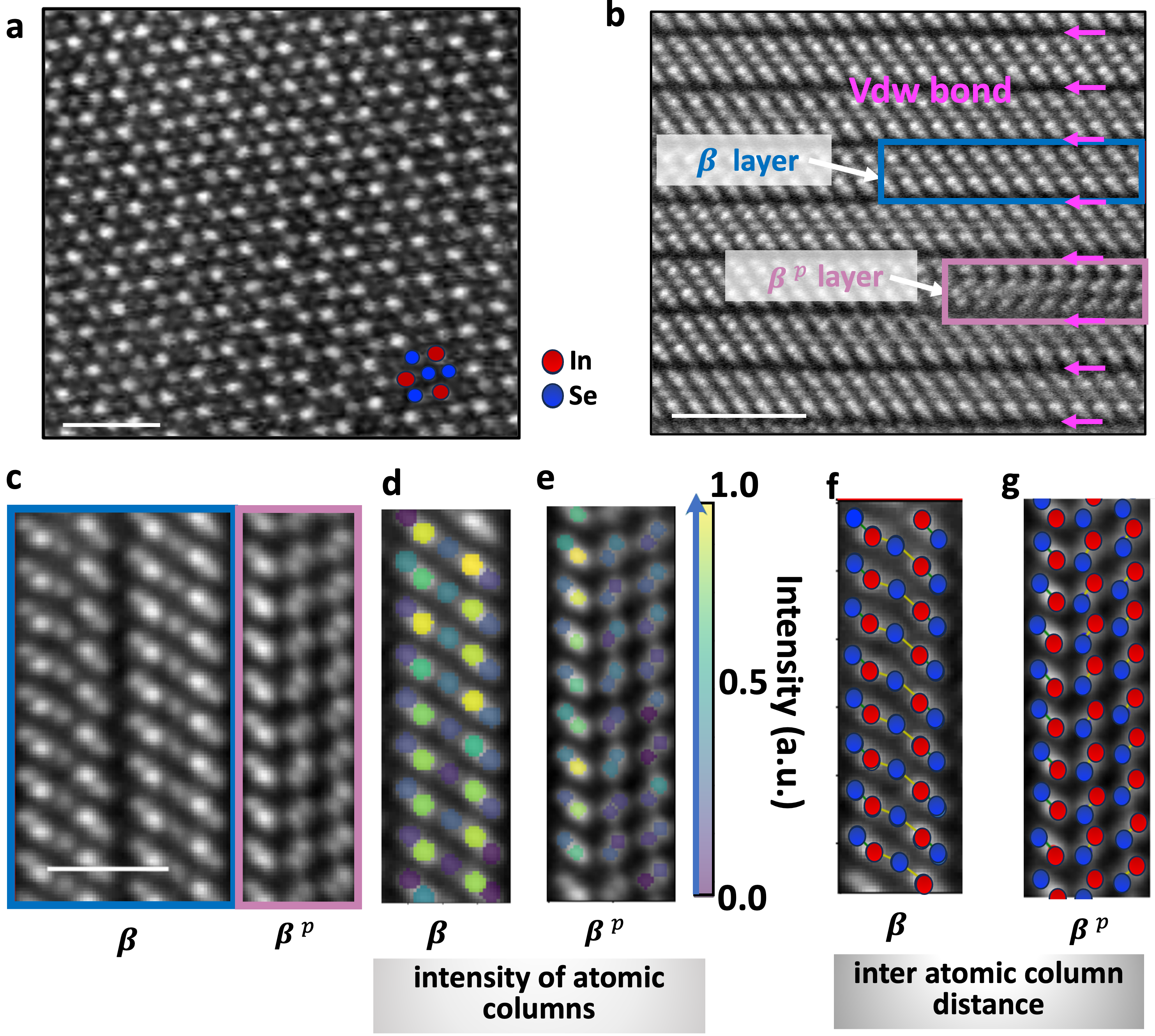}
    \caption{STEM image analysis of the optimized indium selenide films.  
(a) Top-view STEM-ADF image shows that the films primarily consist of the nominally $\upbeta$ phase, scale bar 1 nm. (b) Cross-sectional STEM-ADF image reveals that the film is composed of layers of the $\upbeta$ phase with a 3R polytype configuration. Surprisingly, it also shows unique zigzag atomic layers of indium selenide (designated as the $\upbeta^\text{p}$ phase), scale bar 2 nm. Distinct van der Waals bonds are observed between the phase layers. (c) HAADF-STEM images of the $\upbeta$ phase and the zigzag $\upbeta^\text{p}$ phase, scale bar 2 nm. Sum of intensity for each atomic column in the (d) $\upbeta$ layers, (e) zigzag $\upbeta^\text{p}$ layers. Interatomic column distances in the (f) $\upbeta$ phase, (g) zigzag $\upbeta^\text{p}$ phase. The red and blue circles represent the centers of In and Se atoms, respectively. In the $\upbeta$ phase, the In–Se–In interatomic distance is 0.228 nm (yellow lines), and the Se–In distance is 0.171 nm (green line). In $\upbeta^\text{p}$ phase, the left Se–In interatomic distance is 0.151 nm, and the distance between the two right Se–In atomic columns is 0.132 nm.}
    \label{fig:3}
\end{figure}

\begin{figure}[h]
    \centering
    \includegraphics[width=0.9\textwidth]{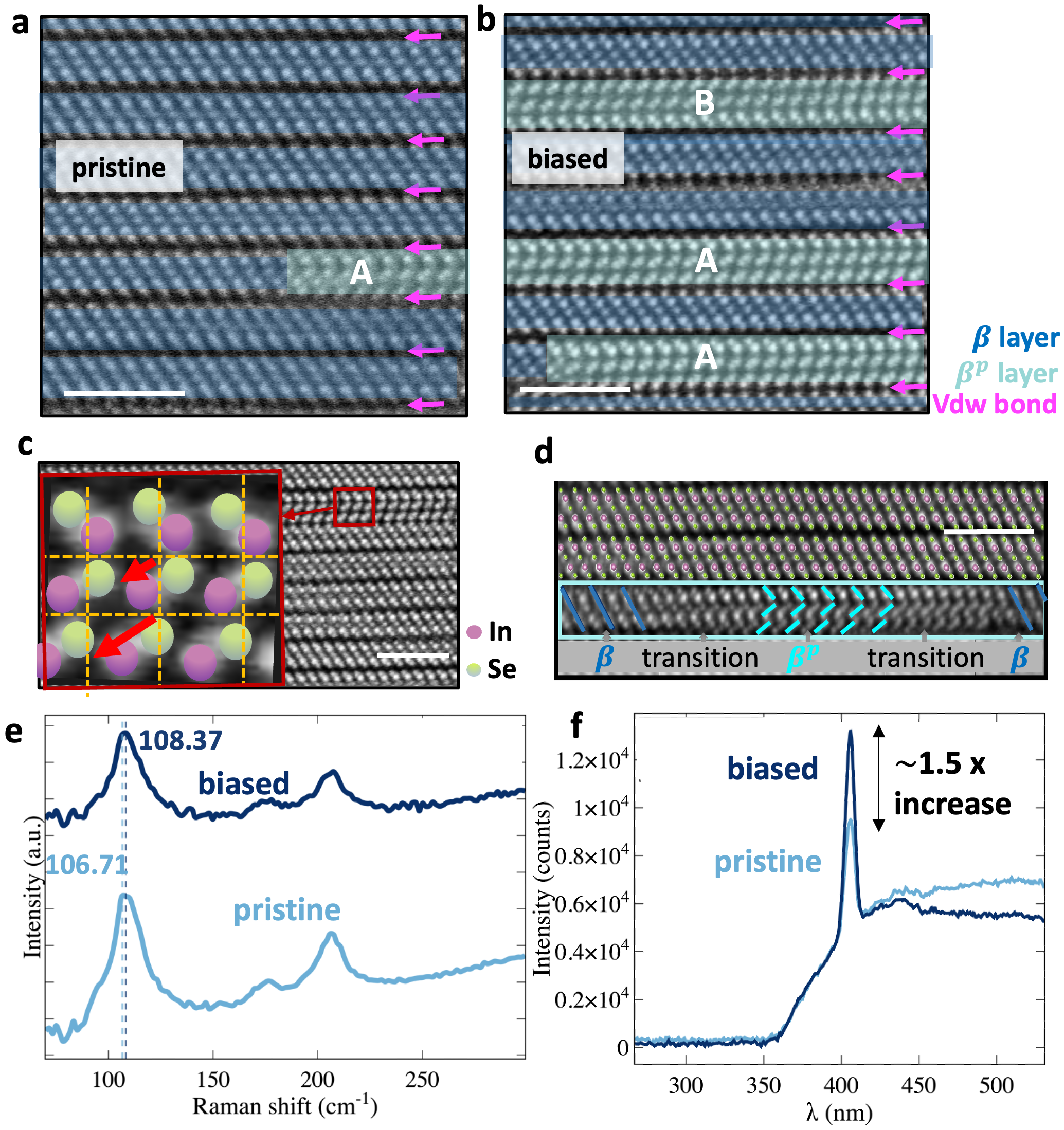}
    \caption{Cross-sectional STEM images of (a) a pristine film with $\upbeta$ phase (blue shaded region) and the $\upbeta^\text{p}$ phase (cyan shaded region); (b) a biased film, demonstrating an increased presence of zigzag $\upbeta^\text{p}$ phases. (c) False-colored STEM image of $\upbeta^\text{p}$ quintuple layer. The $\upbeta^\text{p}$ phase exhibits sliding of the In atoms within the quintuple layer, as highlighted by the red arrow. The centers of the purple and green circles represent In and Se atoms, respectively. This displacement contributes to breaking the local structural mirror symmetry and is therefore the origin of the ferroelectricity. (d) The transition region between the $\upbeta$ and $\upbeta^\text{p}$ phases, scale bar 2 nm. (e) Raman spectra show a slight shift in the Raman peak after bias. (f) Comparison of SHG response in pristine and biased films. The increased proportion of the noncentrosymmetric $\upbeta^\text{p}$ phase under bias results in approximately 1.5-fold enhancement of the SHG signal.}
    \label{fig:4}
\end{figure}

\clearpage

\section{Experimental Section}

\subsection{\label{sec:app:mbe}Synthesis of Indium Selenide Films}

Sapphire substrates (University Wafer; c-plane orientation, $0.2\pm0.1^\circ$ miscut towards the M-plane) were cleaned by sequential rinsing in acetone, methanol, and isopropanol, and were then dried under a nitrogen stream immediately prior to loading into the MBE chamber.

Prior to deposition, all substrates were annealed \textit{in situ} under vacuum at 973 K for at least 10 min, after which the temperature was adjusted to the growth set point. High-purity elemental indium (99.9995~\%, with controlled impurities; Indium Corporation) and selenium (99.999~\%, Alfa Aesar) were co-evaporated from separate effusion cells fitted with pyrolytic boron nitride crucibles. Source fluxes were calibrated at the substrate position using a beam-pressure gauge. At the end of each deposition, all effusion-cell shutters were closed simultaneously and the substrate heater was switched off, allowing the sample to cool to below 400 K. Films were stored under ambient conditions following growth and between all subsequent characterizations.

\subsection{\label{sec:app:xps}X-ray Photoelectron Spectroscopy (XPS)}

XPS measurements were carried out on a Thermo Scientific spectrometer fitted with a monochromatic Al K\(\alpha\) X-ray source (\(h\nu = 1486.6\) eV) and operated at 15 W. High-resolution spectra were collected at a pass energy of 152.6 eV using a 400 \(\upmu\)m diameter analysis spot. To correct for surface charging, all binding energies were referenced to the C 1s hydrocarbon peak at 285.0 eV. Spectral deconvolution employed a Shirley background subtraction and mixed Lorentzian–Gaussian (GL 30) peak shapes. These rigorous acquisition and fitting procedures ensured precise energy calibration and trustworthy surface chemical insights.

\subsection{Experimental Setup and Data Acquisition Procedure for SHG Measurement\label{sec:app:shg}}
 
The SHG measurement was conducted utilizing an 800 nm fundamental light beam generated by a mode-locked Ti: Sapphire pulsed laser (Spectra Physics Solstice ACE). The pulse duration was approximately 89 fs with a repetition rate of 2.5 kHz. Any potential second harmonic component originating from the laser was effectively filtered out using a 425 nm long-pass optical filter. The polarization of the incident light was stabilized in the horizontal direction using a Glan polarizer, while the incident power was regulated via a half-wave plate preceding the polarizer. Further power adjustment was achieved using a neutral density filter. The sample's incident power was quantified using a Germanium-based detector (OPHIR Photonics PD300-IR). The laser beam was precisely focused onto the sample using a 10$\times$ objective lens (Olympus Plan Fluorite), and the resulting second harmonic light at 400 nm was captured utilizing a 20$\times$ objective lens (Zeiss EC Epiplan-Neofluar). Upon transmission through the sample, the laser beam at the fundamental wavelength was isolated through a 600 nm short-pass optical filter. The signal originating from the sample was then directed to a spectrometer (Princeton Instruments IsoPlane SCT 320) for wavelength component separation and was later detected using a CCD detector (Princeton Instruments PIXIS 400). Data were acquired with a 60-second exposure time to ensure an adequate count rate while avoiding sample damage. Finally, the acquired data underwent post-processing using MATLAB\textsuperscript{®} for background subtraction and noise removal.

\subsection{\label{sec:app:xrd}X-Ray Diffraction}

XRD $\uptheta$-2$\uptheta$ measurements were performed in a Rigaku Smartlab XE XRD diffractometer using Cu K$\upalpha$ radiation at a tube current of 50 mA with the X-ray generator running at 40 kV over the selected 2$\uptheta$ range of 5-90$\degree$. The step interval was 0.04$\degree$, with a counting time of 20 s for each step. Scans were performed at a rate of 10$\degree$ per minute.

\subsection{\label{sec:app:raman}Raman Characterization}

Raman spectra were acquired using a Renishaw Qontor™ Dispersive Raman Spectrometer with a 488\,nm excitation laser. Measurements were performed under ambient conditions, with each spectrum collected at a spectral resolution of approximately 1\,cm\(^{-1}\).

\subsection{\label{sec:app:fib}Focused Ion Beam (FIB) Analysis}
Using the \textit{Auto-Slice \& View} feature of the Thermo Fisher Helios 5CX FIB, the sample was milled in one region while an adjacent area was left unmilled for comparison. Each milling cycle removed 50 nm of material in the cross‐sectional direction; after 47 slices, a total depth of
\[
  y = 50\,\text{nm} \times 47 = 2350\,\text{nm}
\]
had been excavated. The Ga$^+$ ion beam was operated at 30\,kV and 2.5\,nA. Imaging was performed with a sub‐nanometer pixel size—ample resolution to detect voids, cracks, or other defects—yet no such imperfections were observed in the fabricated device stack.

\subsection{Electron microscopy characterization and quantitative analysis\label{sec:app:tem}}

Cross-sectional STEM specimens were prepared in a Helios 5CX FIB–SEM, with a final polish at 10 pA and 2 kV. Quantification of the concentration of the $\upbeta^{\mathrm{p}}$ phase was carried out on a Hitachi HD2700 aberration-corrected STEM equipped with a Bruker energy‐dispersive X-ray spectroscopy (EDS) detector at Georgia Institute of Technology, yielding a point‐to‐point resolution of $\sim$0.13 nm. For higher‐resolution measurements of interatomic distances, we used a JEOL NEOARM fifth‐order aberration-corrected STEM (resolution $\sim$0.07 nm) at the University of Pennsylvania. The convergence semi-angle was 25 mrad. The inner and outer semi-angles of the ADF detector were 36 mrad and 127 mrad, respectively. 

Atomic column positions were determined with Atomap \cite{nord2017atomap}, an extension of the HyperSpy framework \cite{francisco_de_la_pena_2025_14956374}. To improve accuracy, refinements of atomic positions were performed by iteratively combining center‐of‐mass analysis with two‐dimensional Gaussian fitting until convergence (deviation $<$ standard deviation). Intensity integration for Figures~\ref{fig:3}(d) and~\ref{fig:3}(e) was performed using the \textit{voronoi} method within radii of 4 and 3 pixels from the atomic‐column centres, respectively. 

Plan‐view specimens (Figure~\ref{fig:3}(a)) were prepared by FIB; the detailed workflow will be described in a forthcoming publication.

\section*{Acknowledgments}
This work was supported by the Air Force Office of Scientific Research MURI entitled, ``Cross-disciplinary Electronic-ionic Research Enabling Biologically Realistic Autonomous Learning (CEREBRAL)" under award number FA9550-18-1-0024. This work was performed, in part, at the Georgia Tech Institute for Electronics and Nanotechnology, a member of the National Nanotechnology Coordinated Infrastructure (NNCI), which was supported by the National Science Foundation (ECCS-1542174). W.C. and A.G. acknowledge support from the NSF under Grant No. DMR-2004749. This work was performed in part at the Singh Center for Nanotechnology at the University of Pennsylvania, a member of the National Nanotechnology Coordinated Infrastructure (NNCI) network, which is supported by the National Science Foundation (Grant NNCI-2025608). We acknowledge Professor Eric Stach and Dr. Jamie Ford at the University of Pennsylvania for their support with the TEM characterization. F.F.A. would like to thank the support from the Cadence Technology Scholarship, the IBM Ph.D. Fellowship, the Stanford Energy Postdoctoral Fellowship, and the Precourt Institute for Energy. Finally, we dedicate this work to the late Professor Evan J. Reed, whose pioneering contributions to the discovery of low-dimensional materials have deeply inspired this study.

\section*{Conflict of Interest}
The authors declare no conflict of interest.

\section*{Data Availability Statement}
The data that support the findings of this study are available from the corresponding author upon reasonable request.

\clearpage

\bibliography{bibfile}

\bibliographystyle{Science}

\clearpage

\section*{Supporting Information}


\setcounter{subsection}{0}
\renewcommand{\thesubsection}{S\arabic{subsection}}

\setcounter{table}{0}
\renewcommand{\thetable}{S\arabic{table}}

\renewcommand{\thefigure}{S\arabic{figure}}
\setcounter{figure}{0}

\begin{figure}[h]
    \begin{subfigure}[t]{\textwidth}
        \centering
\begin{tabularx}{0.9\textwidth}{|Y|Y|Y|Y|Y|}
  \hline
  Indium Flux Ratio (cm$^2$s$^{-1}$)
    & Selenium Flux Ratio (cm$^2$s$^{-1}$)
    & Flux Ratio
    & Substrate Temperature (°C)
    & Deposition Time (min) \\
  \hline
  1.20$\times$10$^{13}$ & 1.97$\times$10$^{14}$ & 16.5 & 700 & 60 \\
  \hline
\end{tabularx}
        \subcaption{}
    \end{subfigure}
    \begin{subfigure}[t]{0.8\textwidth}
    \includegraphics[width=\textwidth]{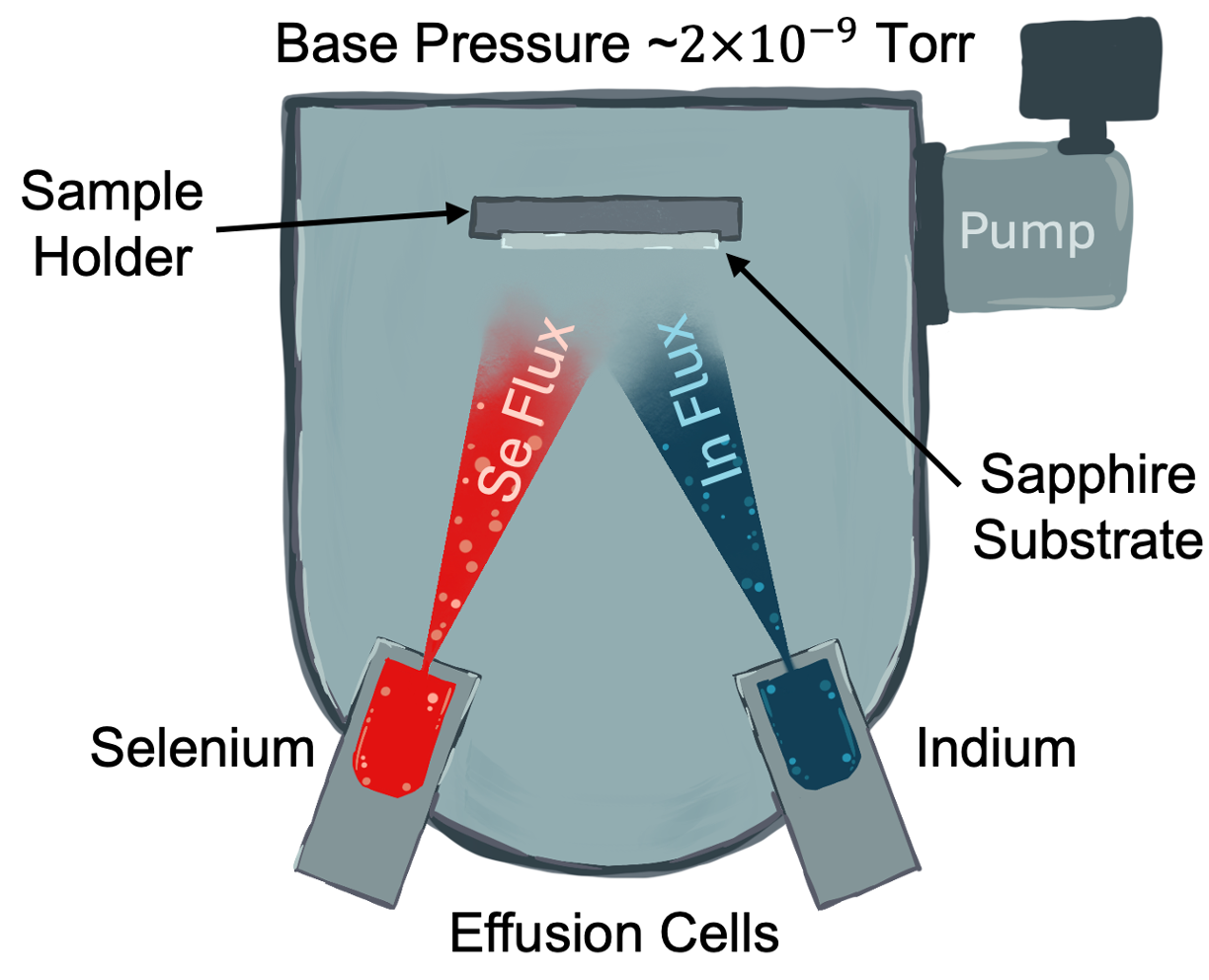}
    \subcaption{}
    \end{subfigure}
    \caption{(a) Detailed synthesis process and growth conditions for large-area optimized indium selenide films deposited on sapphire substrates using MBE. (b) Schematic representation of the MBE setup employed for the deposition, highlighting the key components and configuration.}
    \label{fig:s1}
\end{figure}

\clearpage

\begin{figure}[h]
    \centering
    \includegraphics[width=\textwidth]{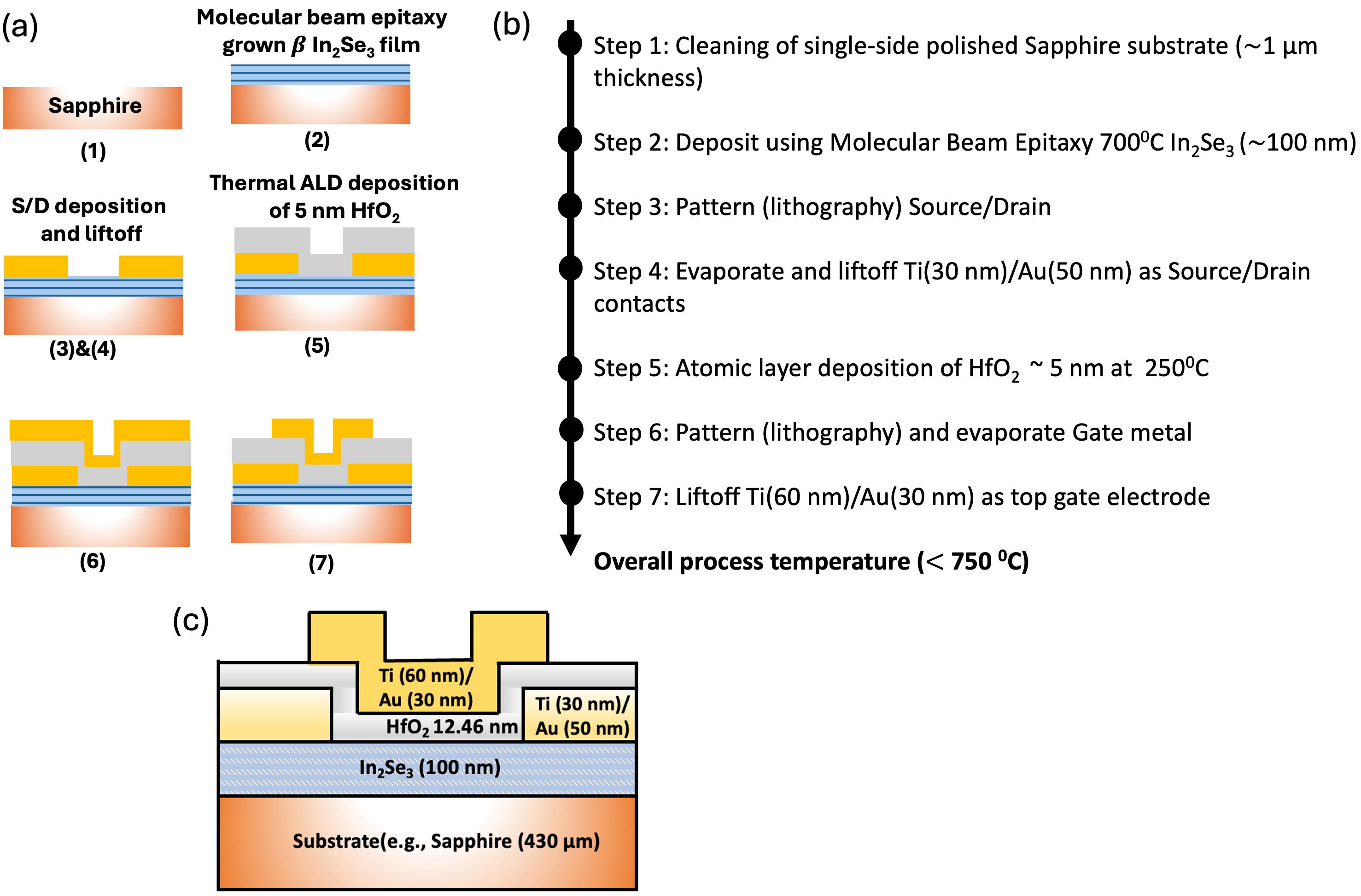}
    \caption{(a) Process flow diagram for the fabrication of the thin-film FET device, detailing each step. (b) Schematic of the fabricated device illustrating the layer dimensions. The entire fabrication process was conducted at temperatures below 700~$\degree$C.}
    \label{fig:s2}
\end{figure}

\begin{figure}[h]
    \centering
    \includegraphics[width=0.5\textwidth]{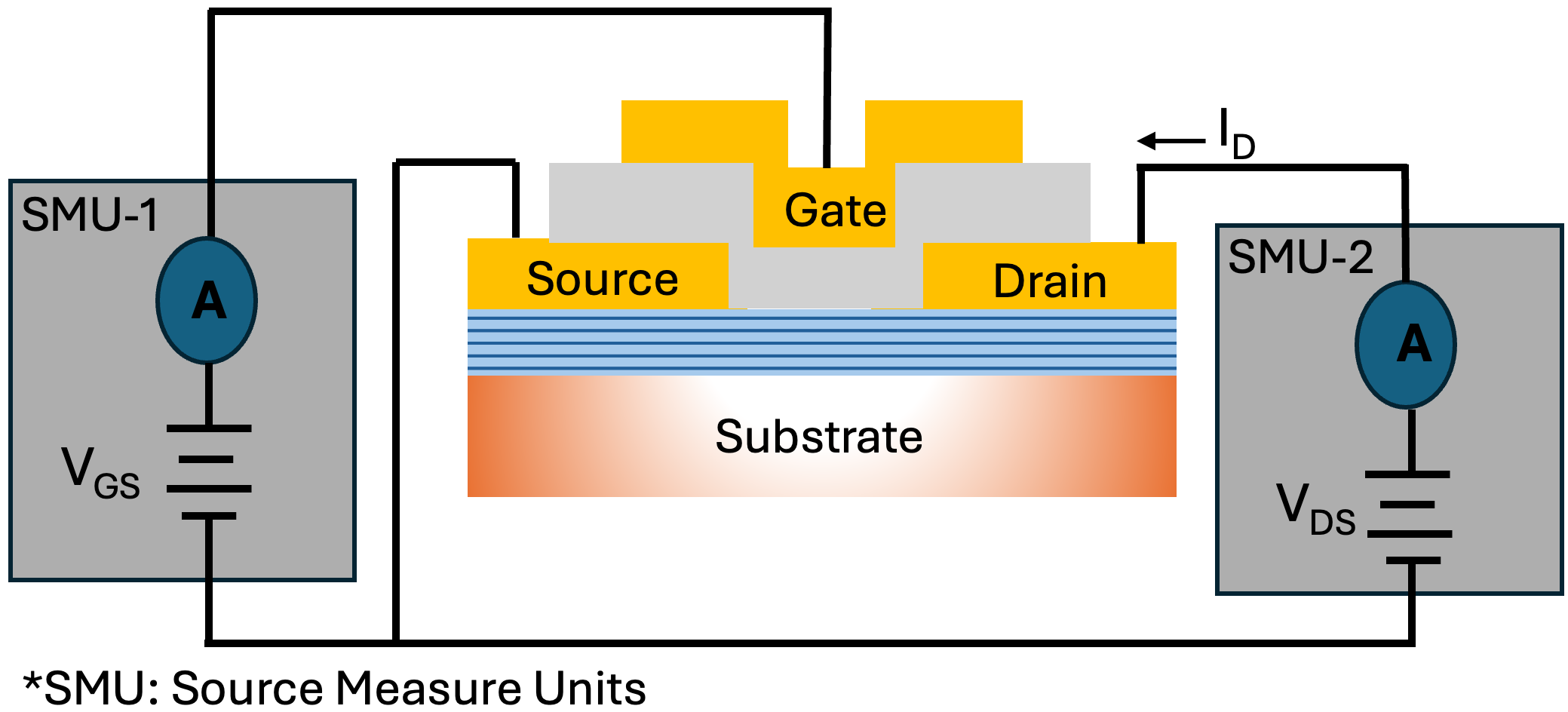}
    \caption{Electrical testing setup and connections. Electrical measurements were performed using two Source Measurement Units (SMUs) as part of a Keithley 4200 Semiconductor Characterization System. A drain-source voltage (V\textsubscript{DS}) was applied between the drain and source terminals, while the gate terminal was independently biased with a gate-source voltage (V\textsubscript{GS}).}
    \label{fig:s3}
\end{figure}

\begin{figure}[h]
    \centering
    \includegraphics[width=0.7\textwidth]{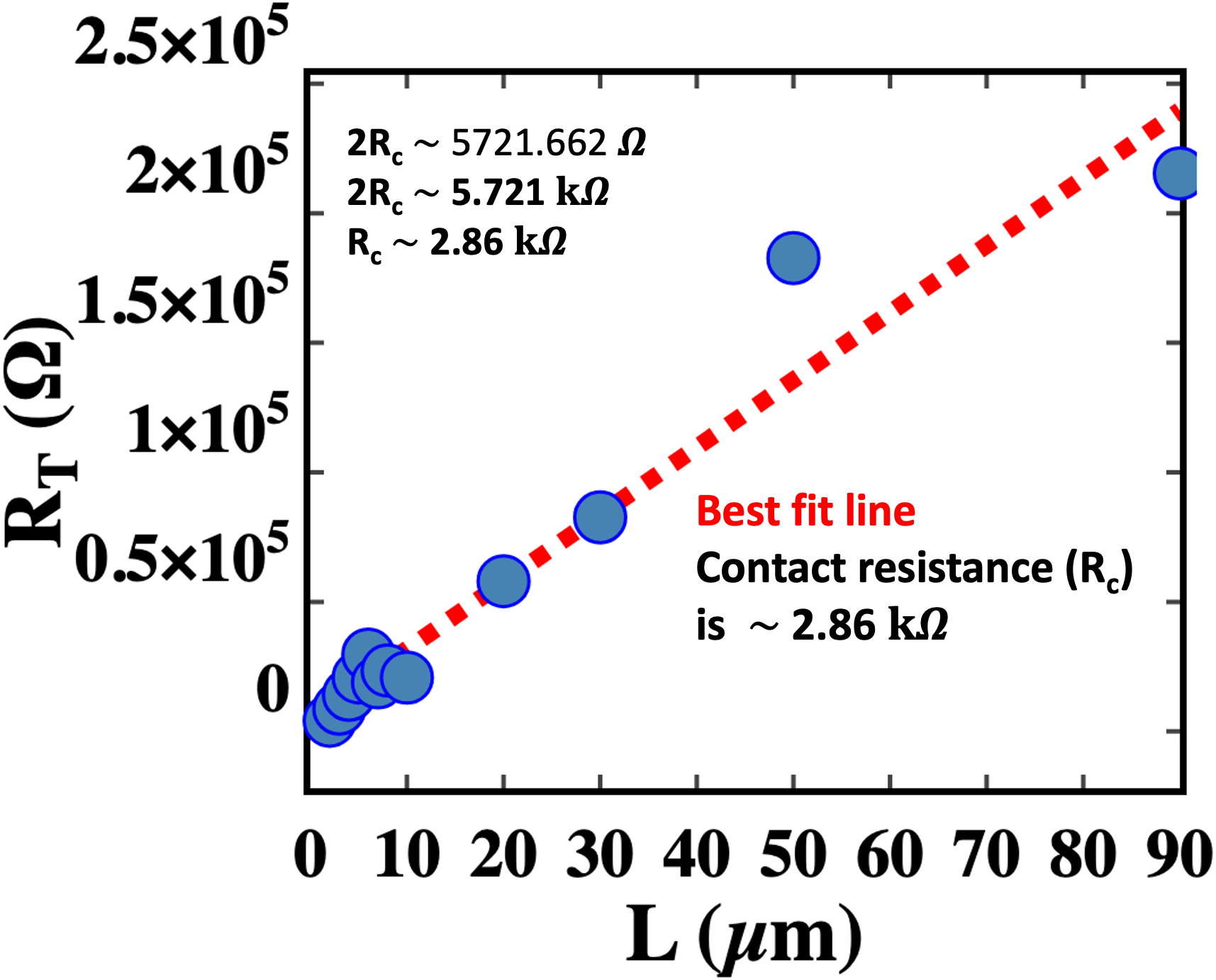}
    \caption{Measurement of R\textsubscript{C} using TLM. The channel lengths used in the measurement ranged from 2 $\upmu$m to 90 $\upmu$m, specifically: 2 $\upmu$m, 3 $\upmu$m, 4 $\upmu$m, 5 $\upmu$m, 6 $\upmu$m, 7 $\upmu$m, 8 $\upmu$m, 10 $\upmu$m, 20 $\upmu$m, 30 $\upmu$m, 50 $\upmu$m, and 90 $\upmu$m, with a constant channel width of 100 $\upmu$m. The applied drain bias (V\textsubscript{DS}) was 2 V, and the total resistance (R\textsubscript{T}) was measured at this bias. From the TLM analysis, the R\textsubscript{C} was determined to be approximately 2.86~k$\Omega$.}
    \label{fig:S4TLM}
\end{figure}

\begin{figure}[h]
    \centering
    \includegraphics[width=\textwidth]{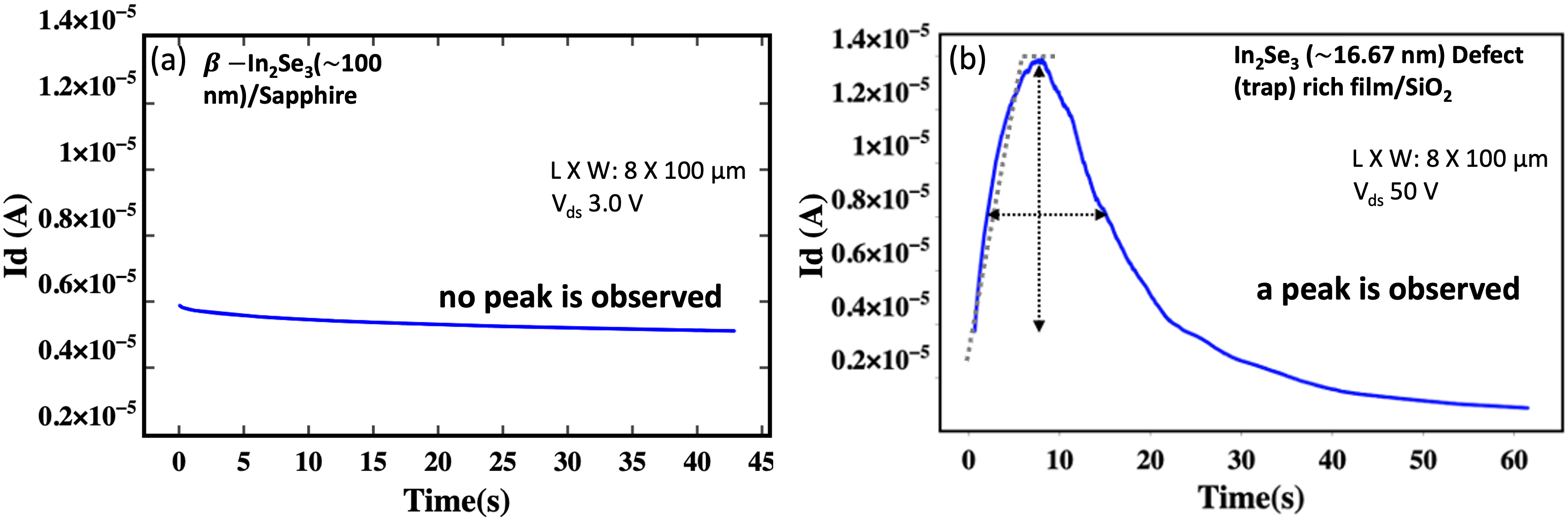}
    \caption{Transient current response analysis conducted on (a) a nominally $\upbeta$-In$_2$Se$_3$ film and (b) a defect-rich In$_2$Se$_3$ film. No peak is observed in the nominally $\upbeta$-In$_2$Se$_3$ film, while a pronounced peak is observed in the trap-rich film, further supporting the conclusion that the observed hysteresis arises from polarization rather than traps.}
    \label{fig:S5}
\end{figure}

\clearpage

\begin{figure}[h]
    \centering
    \includegraphics[width=1\textwidth]{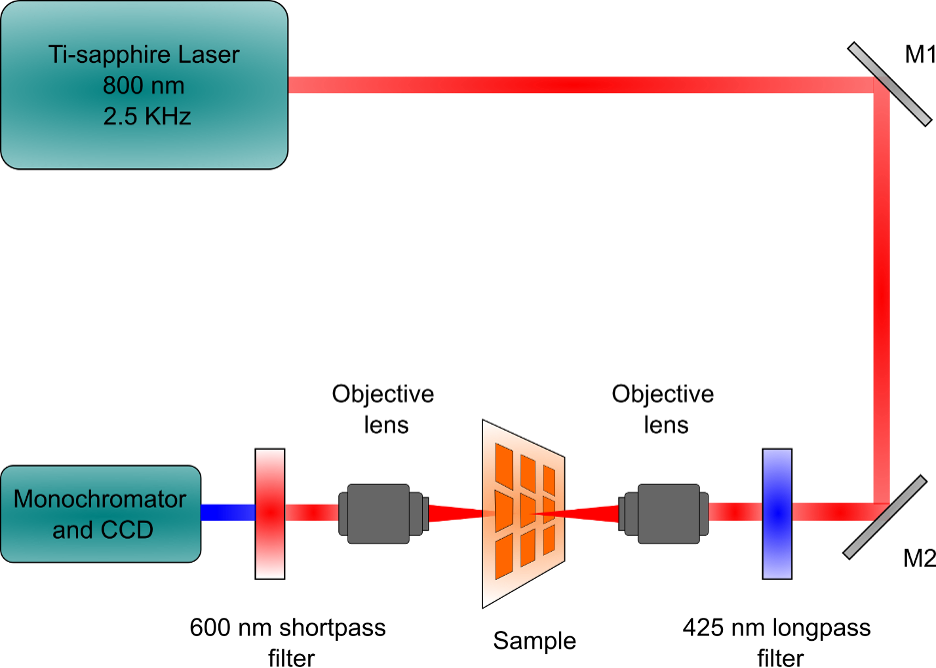}
    \caption{A schematic of the SHG measurement setup. The fundamental light at 800 nm is generated by a Ti: sapphire pulsed laser with a repetition rate of 2.5 kHz. Mirrors M1 and M2 direct the laser light to the sample. A 425 nm longpass filter removes any second harmonic signal originating from the laser itself. The first objective focuses the light on the sample, while the second objective, positioned after the sample, collects the transmitted light. A 600 nm shortpass filter blocks the fundamental laser light, allowing only the second harmonic signal to pass. Finally, a monochromator separates the transmitted light into its spectral components, and a CCD detector records the intensity of each component.}
    \label{fig:s6}
\end{figure}

\begin{figure}[h]
    \centering
    \begin{subfigure}[b]{0.7\textwidth}
        \centering
        \includegraphics[width=\textwidth]{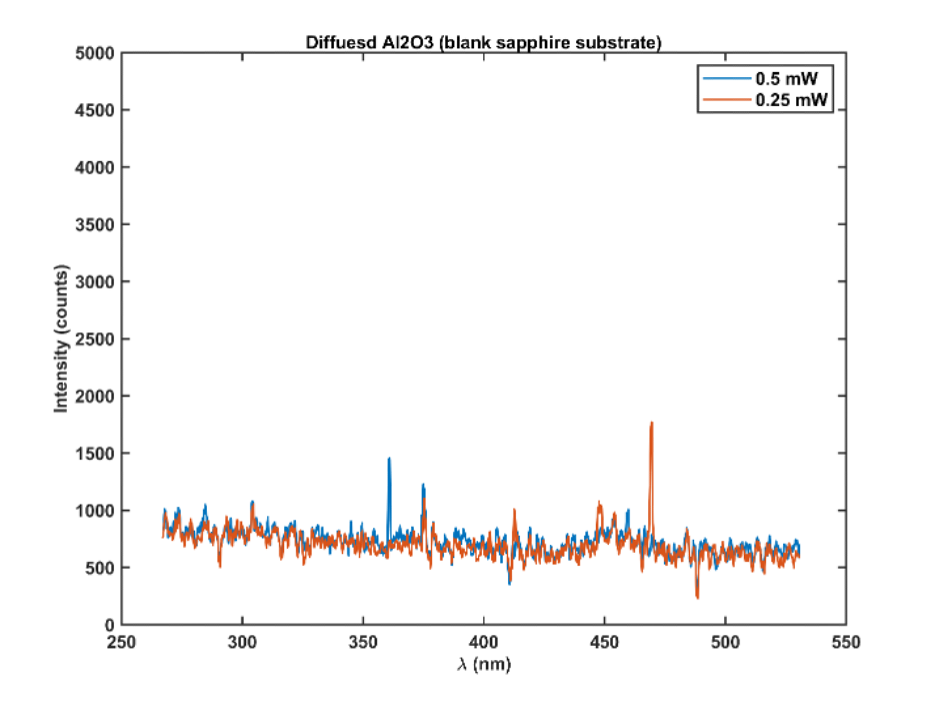}
        \caption{.}
        \label{fig:S7a}
    \end{subfigure}
    \begin{subfigure}[b]{0.7\textwidth}
        \centering
        \includegraphics[width=\textwidth]{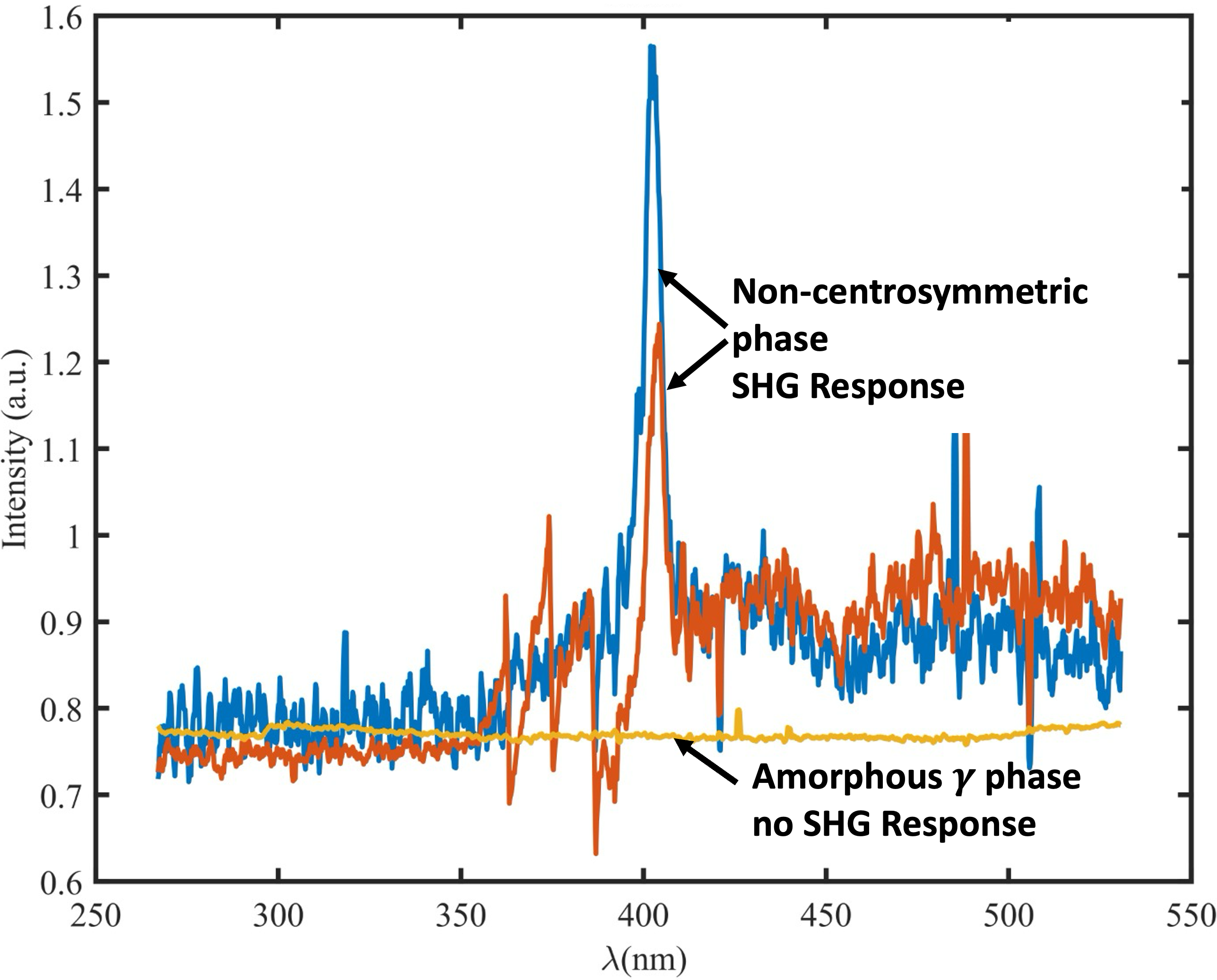}  
        \caption{.}
        \label{fig:S7b}
    \end{subfigure}
    \caption{SHG measurements for (a) blank sapphire substrates and (b) amorphous $\upgamma$-phase In$_2$Se$_3$ films. Both samples exhibited no detectable SHG response, indicating the absence of noncentrosymmetric properties.
}
    \label{fig:S7}
\end{figure}

\clearpage

\begin{figure}[h]
    \centering
    \includegraphics[width=0.9\textwidth]{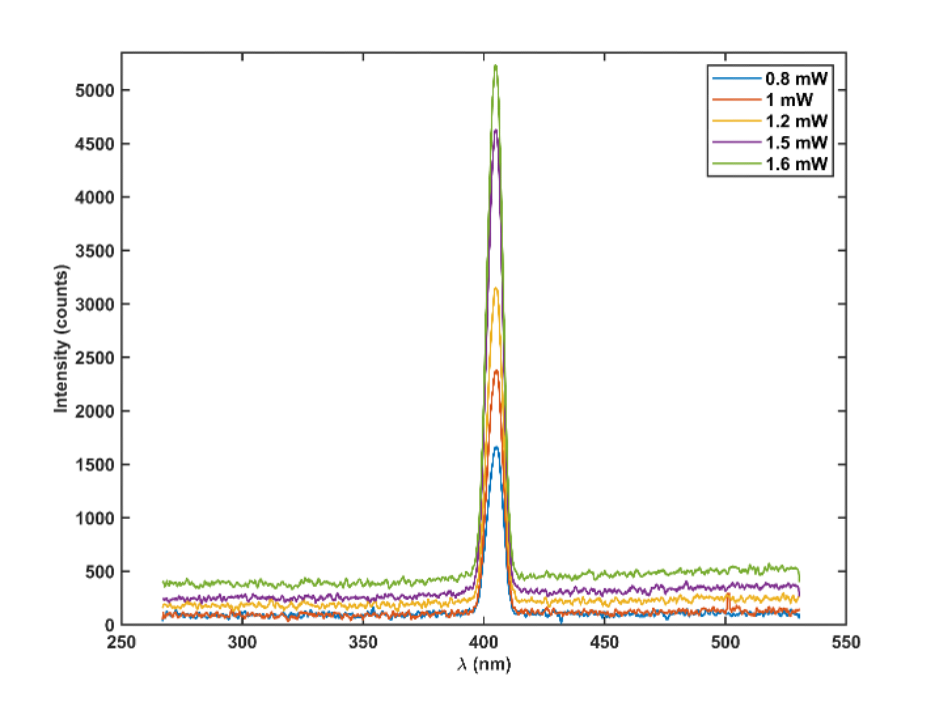}
    \caption{SHG measurements on optimized indium selenide films conducted under varying incident power levels  (0.8, 1.0, 1.2, 1.5, and 1.6 mW) of the fundamental light. These measurements were performed on a distinct batch of films to evaluate the SHG response across different power conditions.
}
    \label{fig:S8}
\end{figure}

\clearpage

\begin{table}[htb]
  \centering
  \begin{tabular}{|cc|cc|c|}
    \hline
    \multicolumn{2}{|c|}{\textbf{Pristine sample}} 
      & \multicolumn{2}{c|}{\textbf{Biased sample}} 
      & \multirow{2}{*}{\shortstack{\textbf{$\upbeta^\text{p}$ region}\\biased vs pristine}} \\
    \cline{1-4}
    \multicolumn{1}{|c|}{\# $\upbeta^\text{p}$ atoms/nm$^2$} 
      & \multicolumn{1}{c|}{$\upbeta^\text{p}$ atomic \%} 
      & \multicolumn{1}{c|}{\# $\upbeta^\text{p}$ atoms/nm$^2$} 
      & \multicolumn{1}{c|}{$\upbeta^\text{p}$ atomic \%} 
      &  \\ 
    \hline
    \multicolumn{1}{|c|}{0.122}   
      & \multicolumn{1}{c|}{0.830\%} 
      & \multicolumn{1}{c|}{2.01}   
      & \multicolumn{1}{c|}{13.6\%} 
      & 16$\times$ \\
    \hline
  \end{tabular}
  \caption{Atomic density and atomic percent of the $\upbeta^\text{p}$ phase region in pristine and biased samples, showing a 16$\times$ increase of the $\upbeta^\text{p}$ phase in the biased sample.}
  \label{tab:1}
\end{table}

\begin{table}[h]
  \centering

  \resizebox{\textwidth}{!}{%
  \setlength{\tabcolsep}{2pt}
  \setlength{\extrarowheight}{9pt}
   \begin{tabular}{|>{\centering\arraybackslash}m{0.8cm}|>{\centering\arraybackslash}m{1.8cm}|>{\centering\arraybackslash}m{1.6cm}|>{\centering\arraybackslash}m{1.8cm}|>{\centering\arraybackslash}m{1.6cm}|>{\centering\arraybackslash}m{0.8cm}|>{\centering\arraybackslash}m{1.8cm}|>{\centering\arraybackslash}m{1.6cm}|>{\centering\arraybackslash}m{1.8cm}|>{\centering\arraybackslash}m{1.6cm}|>{\centering\arraybackslash}m{1.3cm}|}
      \hline
      \multicolumn{5}{|>{\centering\arraybackslash}m{7.6cm}|}{\textbf{Pristine sample}}
        & \multicolumn{5}{>{\centering\arraybackslash}m{7.6cm}|}{\textbf{Biased sample}}
        & \multirow{2}{*}{%
            \shortstack[c]{\\$\upbeta^\text{p}$\\\textbf{region}\\biased\\vs.\\pristine}%
          } \\
      \cline{1-10}
      \shortstack[c]{\# $\upbeta^\text{p}$\\units}
        & \shortstack[c]{\# replaced\\atoms}
        & \shortstack[c]{\\Total\\measured\\(nm$^2$)}
        & \shortstack[c]{\# $\upbeta^\text{p}$\\atoms/nm$^2$}
        & \shortstack[c]{$\upbeta^\text{p}$\\atomic \%}
        & \shortstack[c]{\# $\upbeta^\text{p}$\\units}
        & \shortstack[c]{\# replaced\\atoms}
        & \shortstack[c]{\\Total\\measured\\(nm$^2$)}
        & \shortstack[c]{\# $\upbeta^\text{p}$\\atoms/nm$^2$}
        & \shortstack[c]{$\upbeta^\text{p}$\\atomic \%}
        &  \\
      \hline
      \shortstack[c]{83}
        & \shortstack[c]{415\\(83\,$\times$\,5)}
        & \shortstack[c]{3408}
        & \shortstack[c]{0.122\\\\{\large$\left(\frac{415}{3408}\right)$}}
        & \shortstack[c]{\\0.830\%\\{\large($\frac{0.122}{14.7}$}\\$\times$\,100\%)}
        & \shortstack[c]{1375}
        & \shortstack[c]{6875\\(1375\,$\times$\,5)}
        & \shortstack[c]{3424}
        & \shortstack[c]{2.01\\{\large$\left(\frac{6875}{3424}\right)$}}
        & \shortstack[c]{\\13.6\%\\{\large($\frac{2.01}{14.7}$}\\$\times$\,100\%)}
        & \shortstack[c]{16\,$\times$\\{\large$\left(\frac{13.6}{0.830}\right)$}} \\
      \hline
    \end{tabular}%
  }
  \caption{Detailed quantification of the $\upbeta^\text{p}$ phase region in pristine and biased samples.  
    Columns for each sample show: number of $\upbeta^\text{p}$ units, raw count of replaced atoms, total measured area, computed density, and atomic percentage; the final column gives the ratio of biased vs.\ pristine.}
  \label{tab:S2}
\end{table}

\renewcommand{\arraystretch}{1.0}

\begin{figure}[h]
    \centering
    \includegraphics[width=0.9\textwidth]{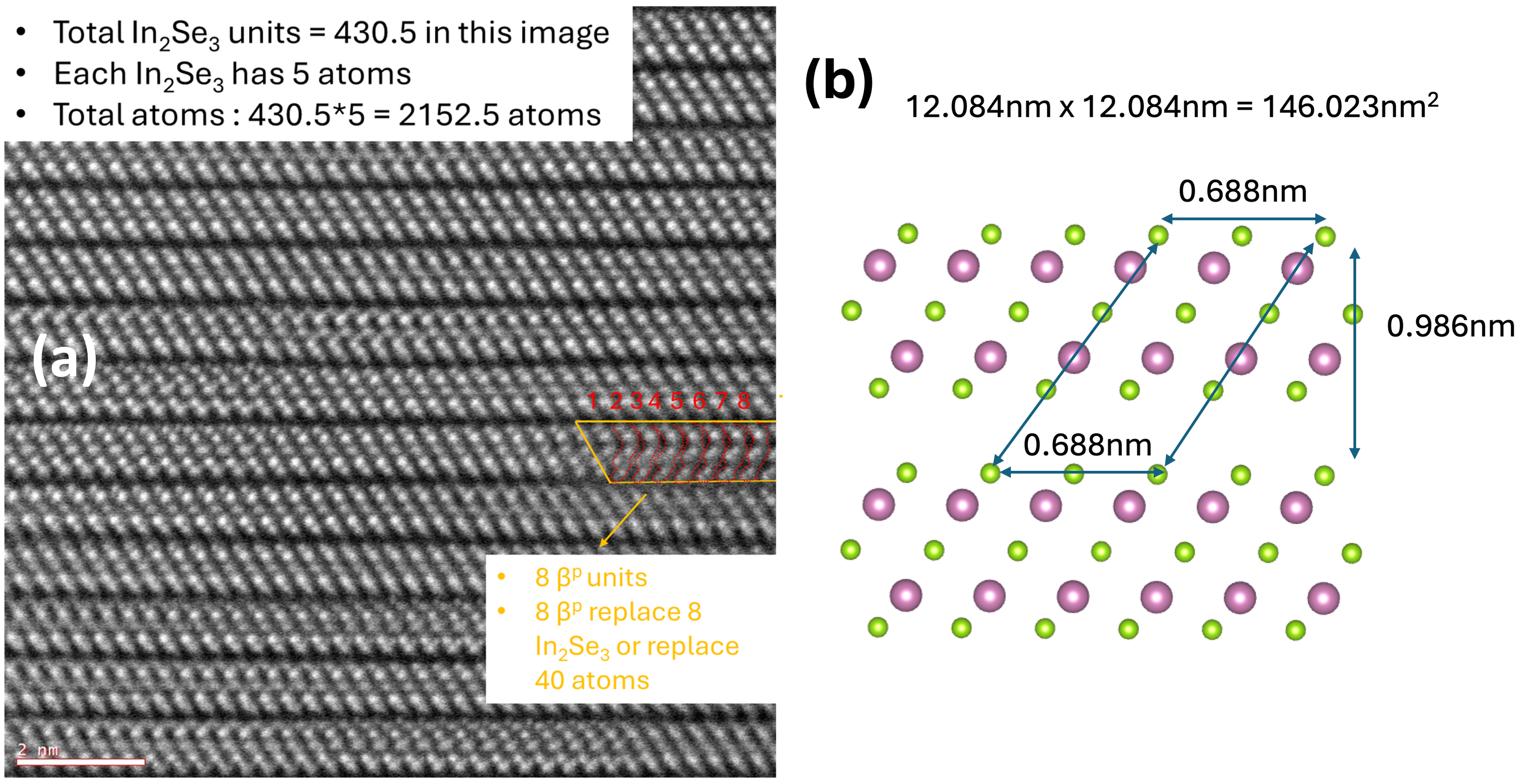}
    \caption{The image dimensions are 12.084 nm $\times$ 12.084 nm = 146.023 nm\textsuperscript{2}. In the parallelogram in b, there are 2 units of In$_2$Se$_3$. The area of the parallelogram is $0.688\text{ nm} \times 0.986\text{ nm} = 0.6784\text{ nm}^2$. Therefore, the areal density of In$_2$Se$_3$ is $\frac{2}{0.6784} = 2.9483~\text{In}_2\text{Se}_3/\text{ nm}^2$, which means there are approximately 2.9483 In$_2$Se$_3$ units per square nanometer. Assuming each unit contains 5 atoms, this corresponds to $2.9483 \times 5 = 14.742\text{ atoms}/\text{nm}^2$. The total number of In$_2$Se$_3$ units in the image is $146.023  \times 2.9483 = 430.5$. Out of these, 8 (inside the orange box) are $\upbeta^\text{p}$ units. Therefore, the concentration of $\upbeta^\text{p}$ configurations in this image is approximately 1.85\% (8/430.5).}
    \label{fig:S9}
\end{figure}

  \begin{figure}[h]
      \centering
      \includegraphics[width=\linewidth]{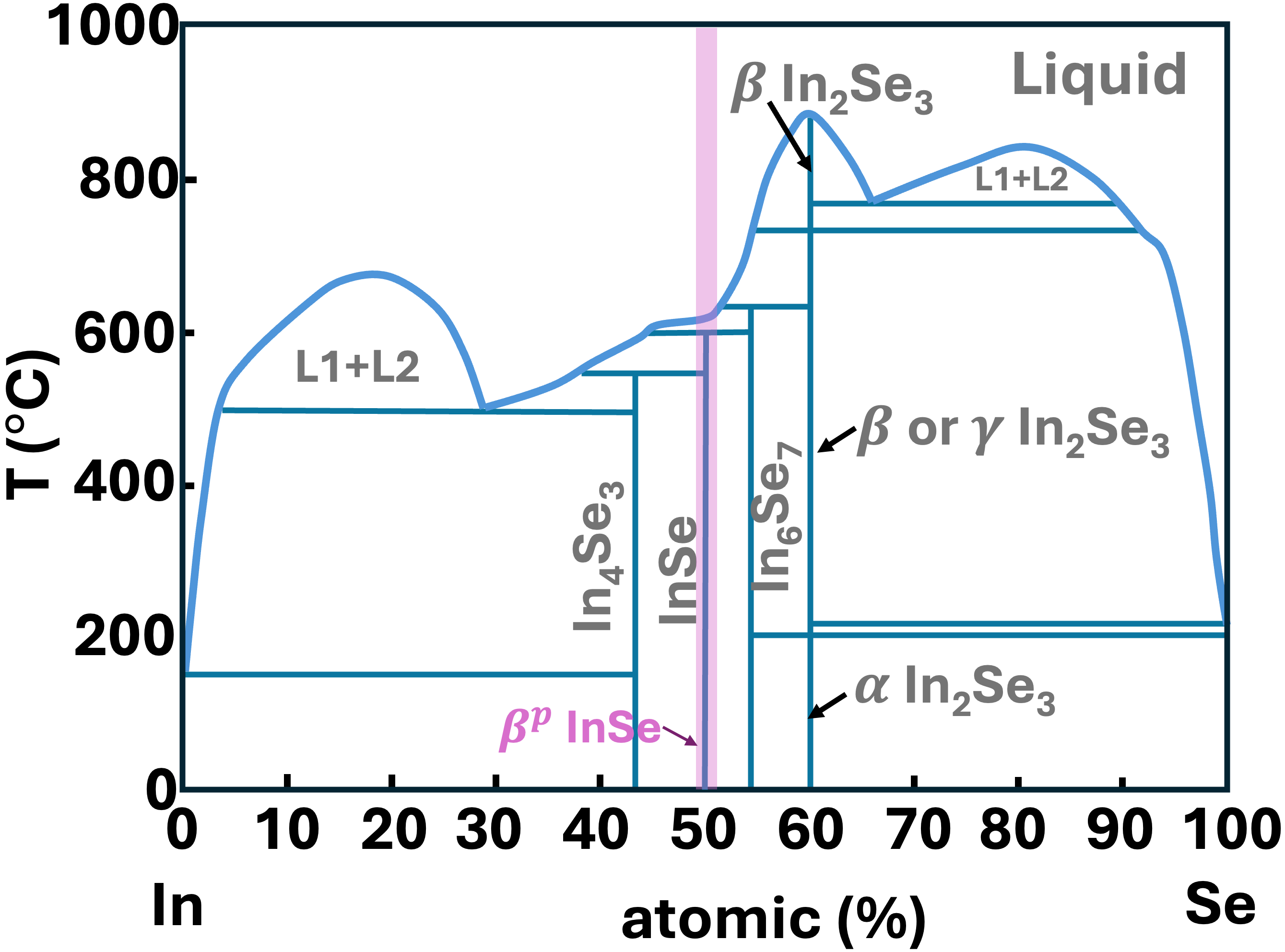}
      \caption{{The phase diagram (adapted from \cite{song2023wafer}) shows the predicted phases at the approximate composition. The $\upbeta^\text{p}$  phase of indium selenide observed at a 50\% atomic ratio (purple shaded region) was previously unreported.}}
      \label{fig:Phase}
  \end{figure}

\clearpage

\begin{table}[ht]
  \centering
  \begin{tabular}{|M{3cm}|M{2cm}|M{1.8cm}|M{6cm}|}
    \hline
    \textbf{Reference} & \textbf{Material} & \textbf{Phase} & \textbf{Stacking Diagram} \\
    \hline
    \textbf{This work} &
       \textbf{InSe} & \textbf{$\upbeta^{\text{p}}$} &
      \includegraphics[valign=c,width=2cm]{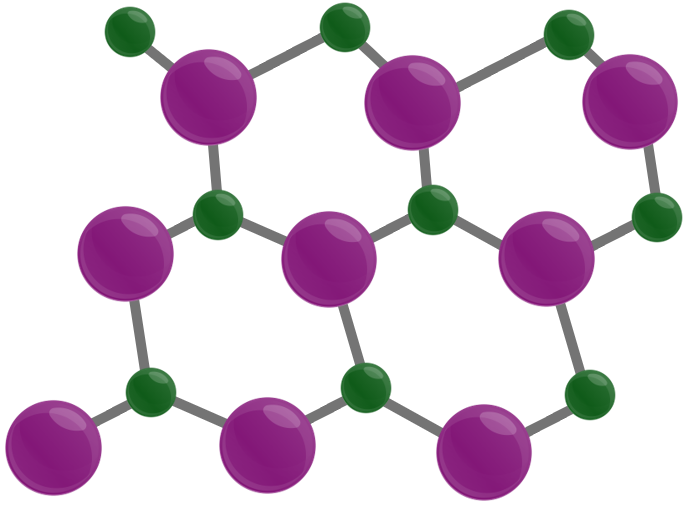} \\
    \hline
    \cite{kupers_controlled_2018} &
         InSe & $\upbeta$ &
      \includegraphics[valign=c,width=1cm]{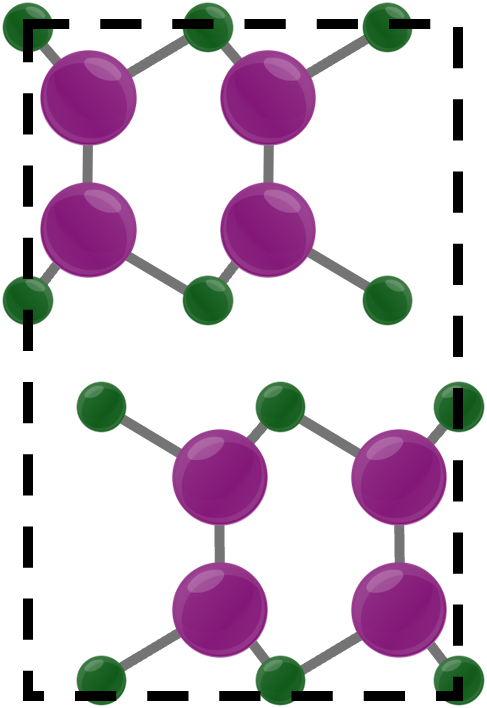} \\
      \hline
      \cite{kupers_controlled_2018} &
         InSe & $\upgamma$ &
          \includegraphics[valign=c,width=1cm]{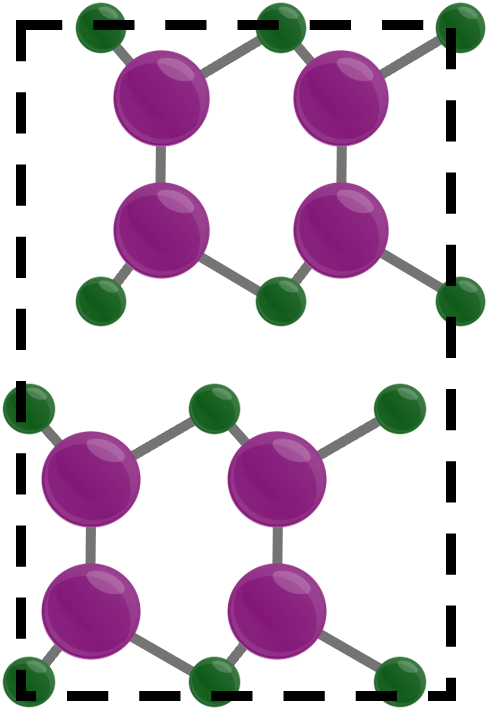} \\
      \hline
      \cite{kupers_controlled_2018} &
         InSe & $\upepsilon$ &
      \includegraphics[valign=c,width=1cm]{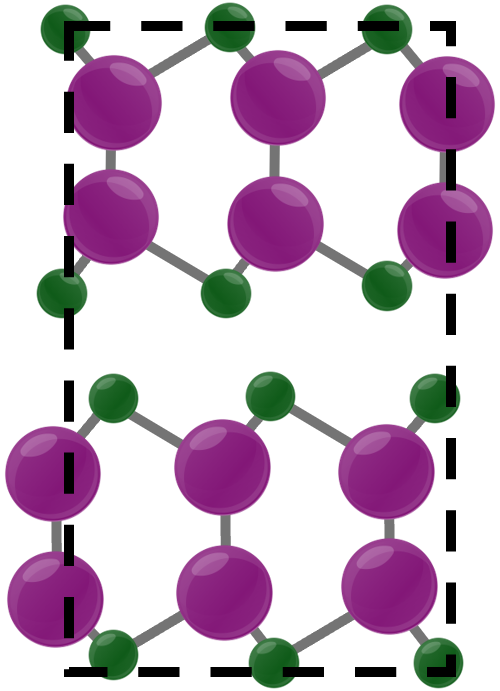} \\
      \hline
    \cite{kupers_controlled_2018}, \cite{fontenele_structural_2023} &
       $\mathrm{In}_{2}\mathrm{Se}_{3}$ & $\upalpha$ - $2H$ &
      \includegraphics[valign=c,width=1.5cm]{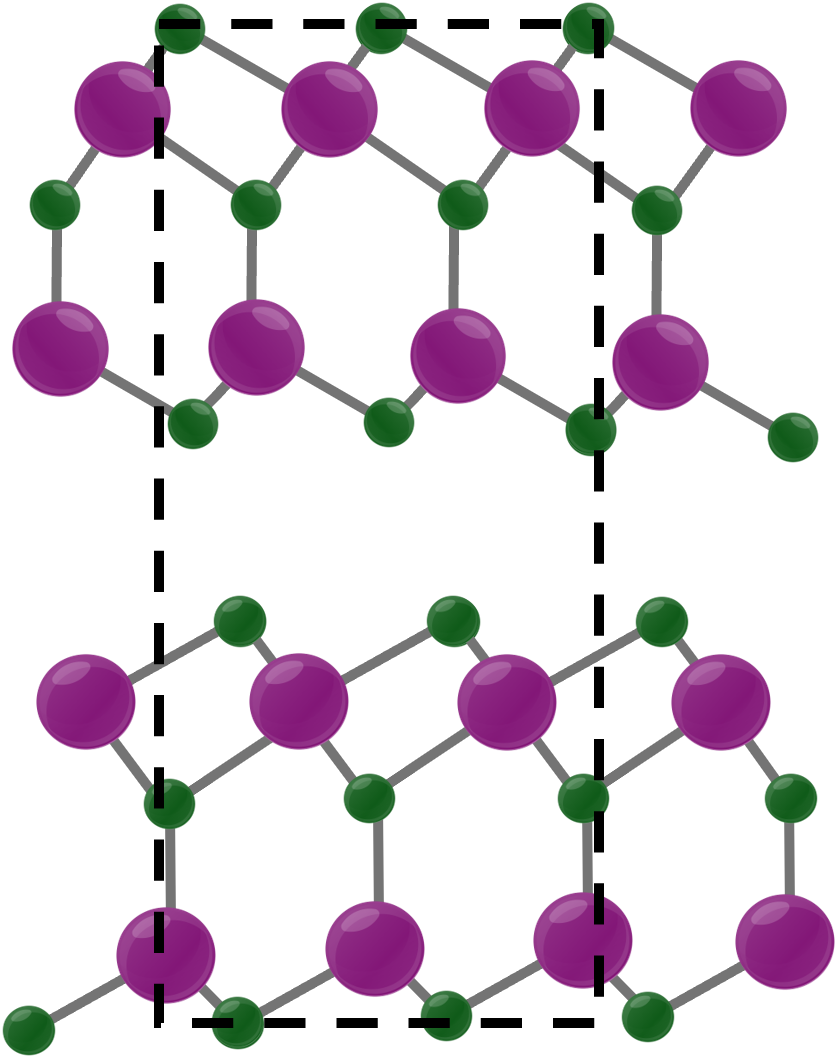} \\
    \hline
    \cite{kupers_controlled_2018}, \cite{fontenele_structural_2023} &
       $\mathrm{In}_{2}\mathrm{Se}_{3}$ & $\upalpha$ - $3R$ &
      \includegraphics[valign=c,width=1.5cm]{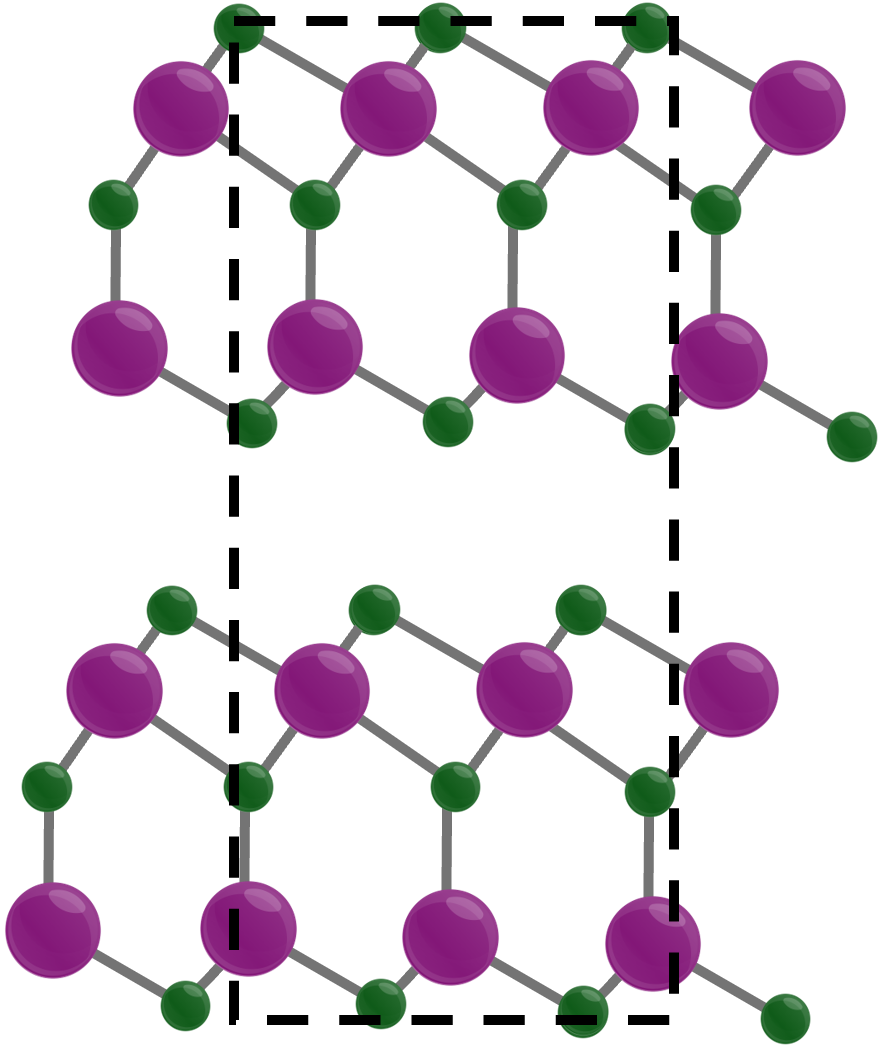} \\
    \hline
    \cite{kupers_controlled_2018}, \cite{fontenele_structural_2023} &
      $\mathrm{In}_{2}\mathrm{Se}_{3}$ & $\upbeta$ - $2H$ &
      \includegraphics[valign=c,width=1.5cm]{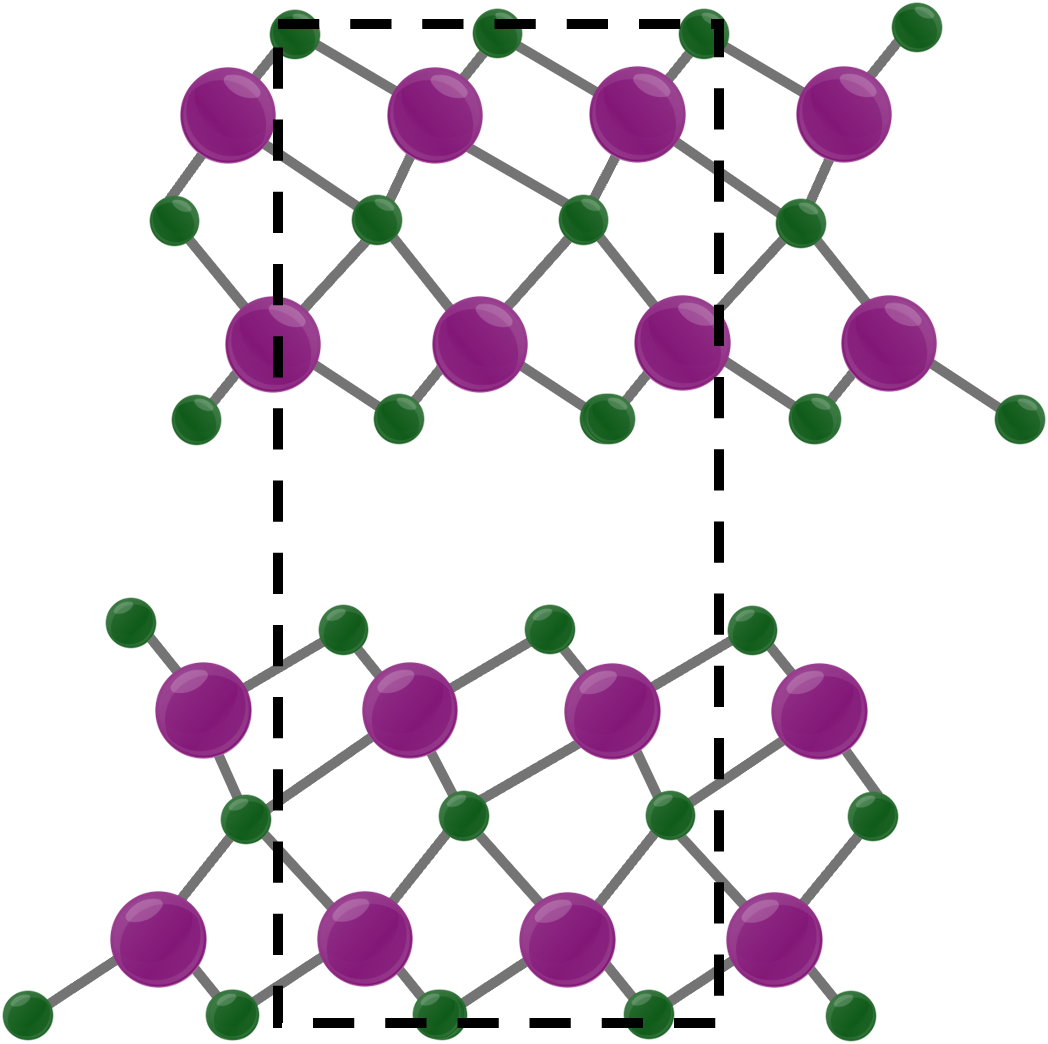} \\
    \hline
    \cite{fontenele_structural_2023} &
       $\mathrm{In}_{2}\mathrm{Se}_{3}$ & $\upbeta$ - $3R$ &
      \includegraphics[valign=c,width=1.5cm]{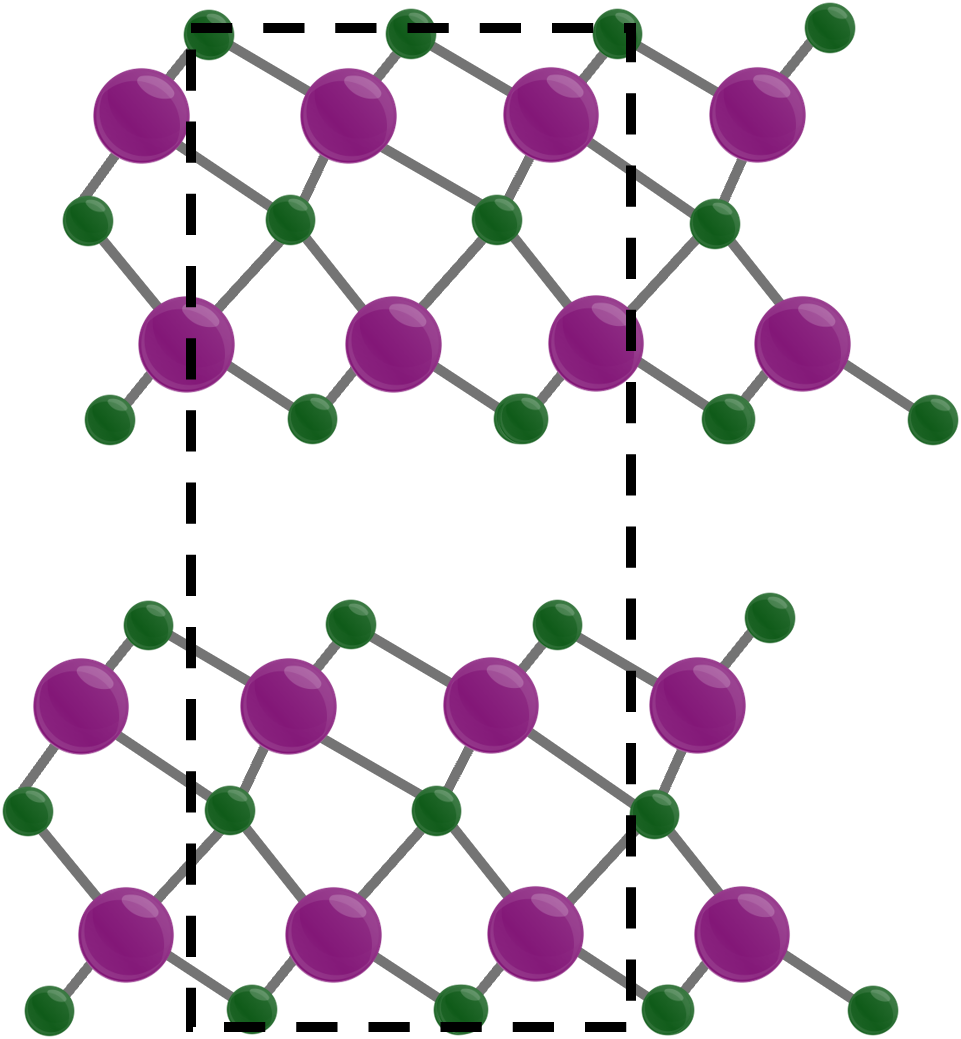} \\
    \hline
    \cite{fontenele_structural_2023} &
       $\mathrm{In}_{2}\mathrm{Se}_{3}$ & $\upbeta$ - $\upbeta'$ &
      \includegraphics[valign=c,width=1.5cm]{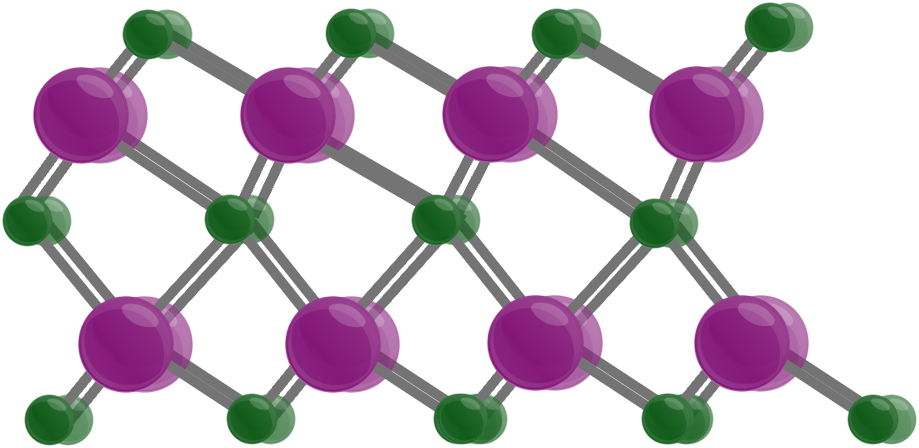} \\
    \hline
    Diagram adapted from Jeengar \emph{et al.}~\cite{jeengar_study_2022} &
       $\mathrm{In}_{2}\mathrm{Se}_{3}$ & $\upgamma$ (3D-structure) &
      \includegraphics[valign=c,height=1cm]{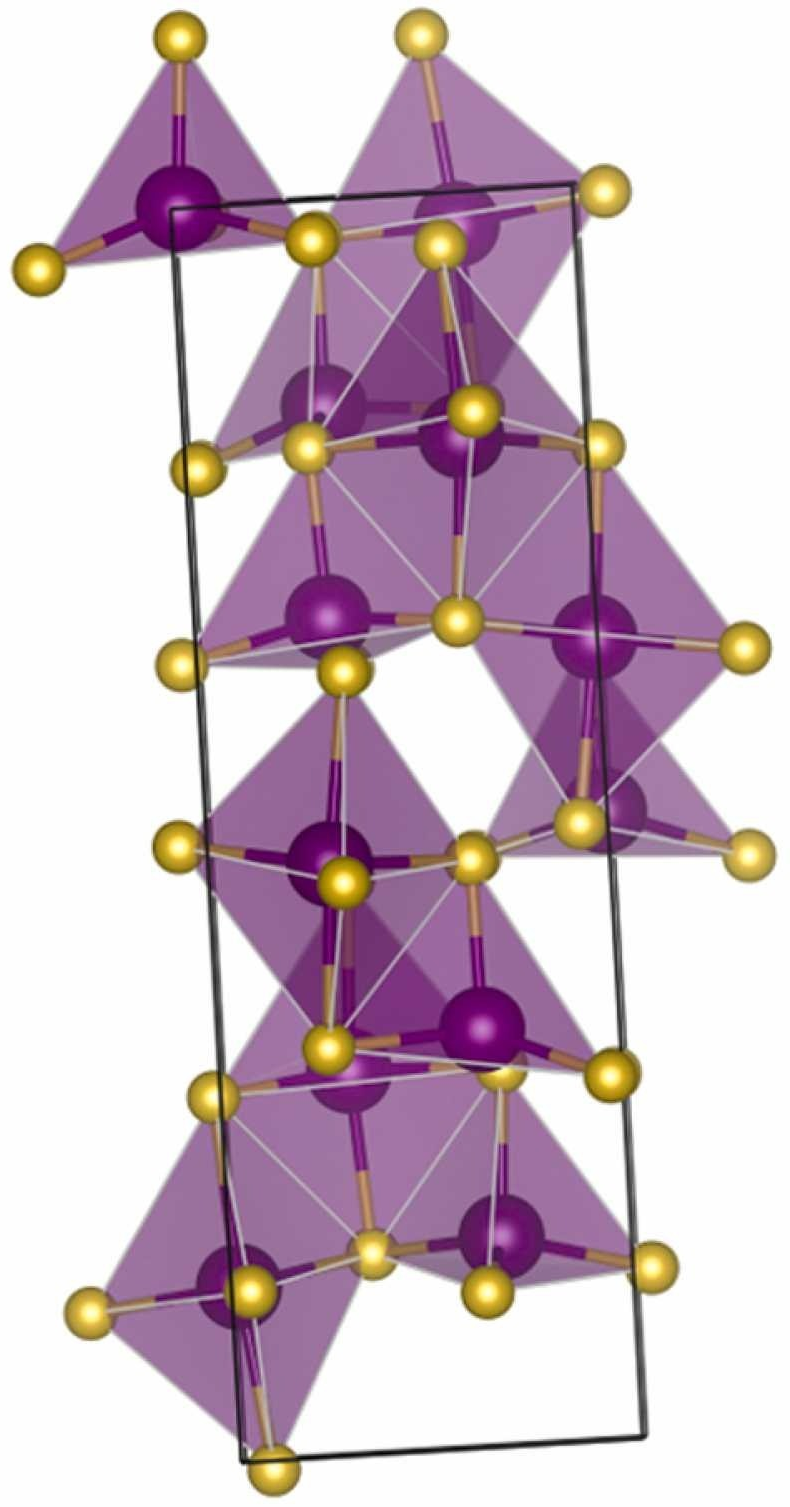} \\
    \hline
    \cite{fontenele_structural_2023} &
       $\mathrm{In}_{2}\mathrm{Se}_{3}$ & $\updelta$ - $2H$ &
      \includegraphics[valign=c,width=1.5cm]{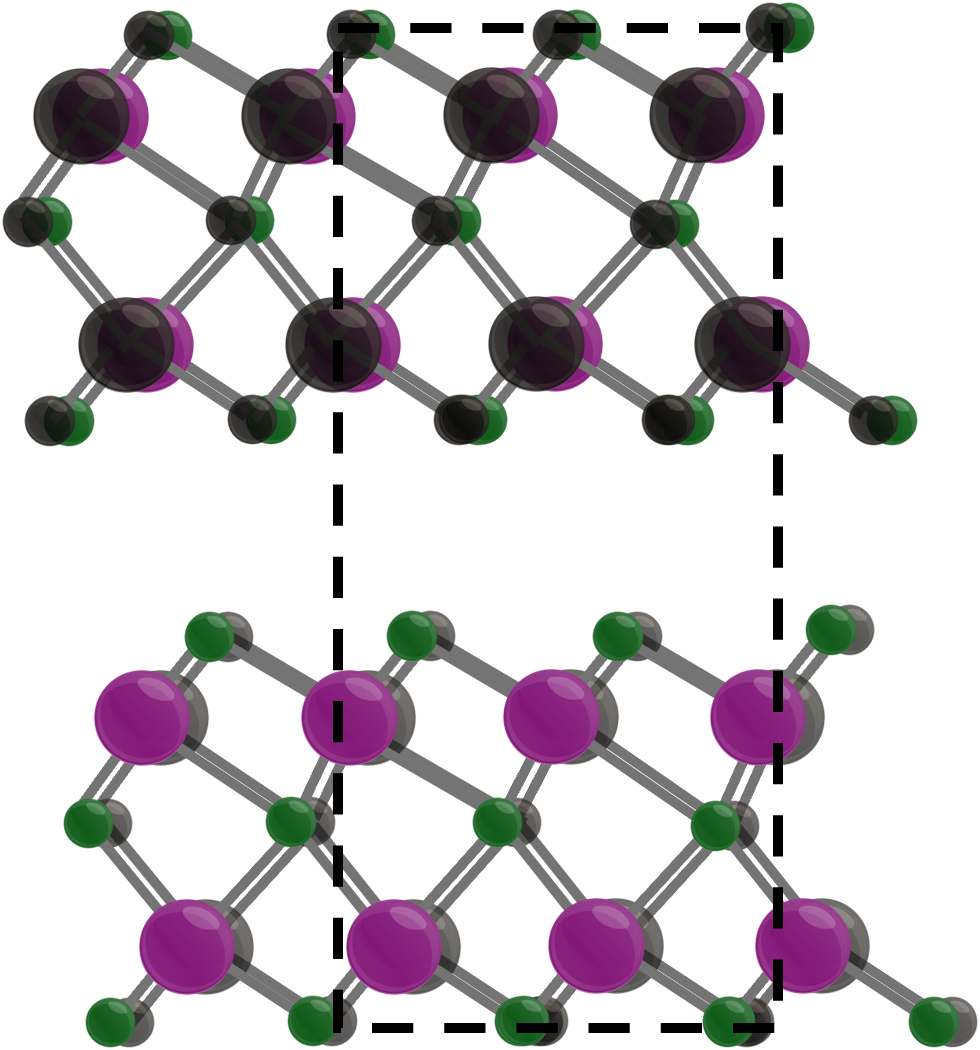} \\
    \hline
    \cite{jasinski_crystal_2002} & 
       $\mathrm{In}_{2}\mathrm{Se}_{3}$ & $\kappa$ &
      No publicly accepted structure, but closely linked to the structure of $\alpha$ \\
    \hline
  \end{tabular}
  \caption{Our synthesized $\upbeta^\text{p}$ phase of indium selenide is distinct from all previously reported phases of InSe and In$_2$Se$_3$. Purple and green atoms represent indium and selenium, respectively. Yellow indicates selenium in the $\upgamma$ structure.}
  \label{tab:phaselist}
\end{table}

\end{document}